\documentclass{jfm}
\usepackage{graphicx}
\usepackage{epstopdf, epsfig}
\usepackage{xcolor}
\usepackage{amsmath}
\usepackage{mathrsfs}

\definecolor{cinnamon}{rgb}{0.82, 0.41, 0.12}
\def\v#1{{\boldsymbol{#1}}}  

\shorttitle{Inertial migration of RBC under a Newtonian fluid}
\shortauthor{N. Takeishi, H. Yamashita, T. Omori, N. Yokoyama, S. Wada, M. Sugihara-Seki}

\title{Inertial migration of red blood cells under a Newtonian fluid in a circular channel}

\author{
  Naoki Takeishi\aff{1} \corresp{\email{takeishi.naoki.es@osaka-u.ac.jp}},
  Hiroshi Yamashita\aff{2},
  Toshihiro Omori\aff{3},\\
  Naoto Yokoyama\aff{4},
  Shigeo Wada\aff{1}, \and
  Masako Sugihara-Seki\aff{1,5}
}

\affiliation{
  \aff{1} Graduate School of Engineering Science, Osaka University, 1-3 Machikaneyama, Toyonaka, Osaka 560-8531, Japan
  \aff{2} Graduate School of Integrated Sciences for Life, Hiroshima University, 1-3-2 Kagamiyama, Higashi-Hiroshima, Hiroshima, 739-8511, Japan.
  \aff{3} Department of Finemechanics, Tohoku University, 6-6-01 Aoba, Sendai, Miyagi, 980-8579, Japan.
  \aff{4} Department of Mechanical Engineering, Tokyo Denki University, 5 Senju-Asahi, Adachi, Tokyo, 120-8551, Japan.
  \aff{5} Kansai University, Department of Pure and Applied Physics, 3-3-35 Yamate-cho, Suita, Osaka 564-8680, Japan
}

\date{First submission 7th Jun. 2022; revised \today}
\begin{document}
\maketitle

\begin{abstract}
We present a numerical analysis of the lateral movement and equilibrium radial positions of red blood cells (RBCs) with major diameter of 8 $\mu$m under a Newtonian fluid in a circular channel with 50-$\mu$m diameter.
Each RBC,
modelled as a biconcave capsule whose membrane satisfies strain-hardening characteristics,
is simulated for different Reynolds numbers $Re$ and capillary numbers $Ca$, 
the latter of which indicate the ratio of the fluid viscous force to the membrane elastic force.
The effects of initial orientation angles and positions on the equilibrium radial position of an RBC centroid are also investigated.
The numerical results show that depending on their initial orientations,
RBCs have bistable flow modes,
so-called rolling and tumbling motions.
Most RBCs have a rolling motion.
These stable modes are accompanied by different equilibrium radial positions,
where tumbling RBCs are further away from the channel axis than rolling ones.
The inertial migration of RBCs is achieved by alternating orientation angles,
which are primarily affected by the initial orientation angles.
Then the RBCs assume the aforementioned bistable modes during the migration,
followed by further migration to the equilibrium radial position at much longer time periods.
The power (or energy dissipation) associated with membrane deformations is introduced to quantify the state of membrane loads.
The energy expenditures rely on stable flow modes,
the equilibrium radial position of RBC centroids,
and the viscosity ratio between the internal and external fluids.
\end{abstract}

\begin{keywords}
  red blood cell,
  inertial migration,
  channel flow,
  capsule,
  computational biomechanics.
\end{keywords}

\section{Introduction}
Particle migration at a finite channel (or particle) Reynolds number $Re$ ($Re_\mathrm{p}$) in microchannels has been intensively studied not only from the viewpoint of pure physics,
but also in terms of bioengineering applications such as label-free cell alignment, sorting, and separation techniques~\citep{Martel2014, Warkiani2016, Zhou2019}. 
Such microfluidic techniques allow us to reduce the complexity and costs of clinical applications by using small amount of blood samples.
While a number of studies have analysed the inertial migration of rigid spherical particles using a variety of approaches,
such as analytics~\citep{Asmolov1999, Ho1974, Schonberg1989},
numerical simulations~\citep{Bazaz2020, Feng1994, Yang2005},
and experimental observations~\citep{Carlo2009, Karnis1966, Matas2004},
the inertial migration of biological cells,
which can be assumed to be deformable particles consisting of the internal fluid enclosed by a thin membrane,
has not yet been fully described.
Red blood cells (RBCs) are a major component of human blood cells,
with a volume fraction of 45\% (the other 55\% is plasma) and a density of $\sim$5 million/mL.
The behaviour of individual RBCs subject to finite $Re$ is of paramount importance in manipulating cells or quantifying cell state.
Due to their unique biconcave shape and high deformability,
it is expected that the problem of inertial migration of RBCs is made more complex in comparison with rigid spherical particles originally reported by~\cite{Segre1962},
where the particles exhibit lateral movement and flow in the equilibrium position away from the channel center as a consequence of the force balance between the shear-induced and wall-induced lift forces,
the so-called ``inertial migration'' or ``tubular pinch effect''~\citep{Segre1962}.
\cite{Jeffery1922} speculated that an ellipsoid may alter its orientation so that the viscous energy dissipation of the system becomes minimal.
However, this is not true for soft particles with large deformation.
Although many former studies have examined the dynamics of a non-spherical capsule, e.g., in~\cite{Omori2012},
none of them have fully answered this question.
The behaviour of a single, almost inertialess RBC in a microchannel whose scale is comparable to the cell size has been well investigated,
e.g., in~\cite{Fedosov2014, Guckenberger2018, Noguchi2005, Takeishi2021}.
These studies have revealed velocity-dependent transitions in RBC shapes.
\cite{Tomaiuolo2009} found parachutes at smaller velocity ($\sim$ 0.11 cm/s) and slippers at higher velocity ($\sim$ 3.6 cm/s) in circular channels of 10-$\mu$m diameter.
\cite{Cluitmans2014} detected croissants at lower velocities ($\leq$ 5 mm/s) and slippers at higher velocities ($\geq$ 10 mm/s) in square channels with widths $\leq$ 10 $\mu$m.
Using the same parameters as~\cite{Cluitmans2014},
\cite{Quint2017} found a stable slipper and a metastable croissant in a rectangular channel of 25 $\mu$m $\times$ 10 $\mu$m.
The shape transition from croissant/parachute to slipper shape was also identified in more recent study by~\cite{Guckenberger2018} that used a rectangular channel of 12 $\mu$m $\times$ 10 $\mu$m.
The slipper shape was associated with an off-centered position~\citep{Guckenberger2018},
which is counter to traditional knowledge about the axial focusing of spherical deformable particles toward the channel axis~\citep{Karnis1963}.
Hereafter, we call this phenomenon as ``axial migration''.
A more recent numerical study further showed that compared to the parachute shape,
the off-center slipper shape had low energy expenditure associated with membrane deformations,
and the equilibrium radial positions of these two RBC centroids correlated well with the energy expenditure~\citep{Takeishi2021}.
Despite of these insights,
it is still uncertain whether the aforementioned stable shapes of RBCs persist even under finite inertia in larger microchannels with diameters of several dozen micrometers,
and whether the equilibrium radial positions can be described by the energy expenditure.
Therefore, the objective of this study was to clarify the relationship between the stable flow mode of RBCs,
equilibrium radial position,
and energy expenditure associated with membrane deformations in several dozen circular microchannels for finite $Re$.

So far, inertial migration of rigid spherical particles have been well investigated, e.g., in~\cite{Martel2014, Morita2017, Nakagawa2015, Nakayama2019}.
Under moderate $Re_\mathrm{p}$,
the rigid particles align in an annulus at a radius of about 0.6$R$,
where $R$ is the channel wall radius~\citep{Segre1962}.
The radius of the equilibrium annulus increases with $Re_\mathrm{p}$ because of the increase in the shear-induced inertial lift force~\citep{Matas2004, Matas2009}.
The equilibrium position 0.6$R$ was observed for $Re= 2R\overline{V}/\nu = O(1)$ and shifted to larger radius for larger $Re$,
where $\overline{V}$ was the average axial velocity~\citep{Matas2004}.
Studies of inertial migration of biological cells have attracted particular attentions recently~\citep{Warkiani2016, Zhou2019}.
For instance, \cite{Hur2011} experimentally investigated the inertial migration of various cell types (including RBCs, leukocytes, and cancer cells such as a cervical carcinoma cell line, breast carcinoma cell line, and osteosarcoma cell line) with a cell-to-channel size ratio 0.1 $\leq d/W \leq$ 0.8,
using a rectangular channel with a high aspect ratio of $W/H \approx$ 0.5,
where $d$, $W$ and $H$ are the cell diameter, channel width, and height, respectively.
Their results showed that the cells could be separated according to their size and (elastic) deformability~\citep{Hur2011}.
The experimental results can be qualitatively described by a spherical capsule model~\citep{Kilimnik2011} and droplet model~\citep{Chen2014}.
In a more recent experiment by~\cite{Hadikhani2018} involving bubbles in rectangular microchannels and different bubble-to-channel size ratios 0.48 $\leq d/W \leq$ 0.84,
the authors investigated the effect of bubble diameter $Re$ (1 $< Re <$ 40) and capillary number $Ca$ (0.1 $< Ca <$ 1) on the lateral equilibrium,
where $Ca$ is the ratio between the fluid viscous force and the membrane elastic force.
The equilibrium position of such soft particles results from the competition between $Re$ and $Ca$,
because at high $Re$,
the flow pushes the particles towards the wall,
while at high $Ca$, i.e., high deformability, particles can move towards the channel center.
Numerical analysis more clearly showed that ``deformation-induced lift force'' became stronger as the particle deformation increased~\citep{Raffiee2017, Schaaf2017}.
Although numerical analysis of inertia migration has been intensively investigated in recent years mostly for spherical particles~\citep{Bazaz2020},
equilibrium positions of soft particles is still debated owing to the complexity of the phenomenon.
\cite{Shin2011} investigated the equilibrium position of a two-dimensional spherical capsule in the range of 1 $\leq Re \leq$ 100 for different capsule-to-channel size ratios 0.1 $\leq d/H \leq$ 0.4.
Their numerical results showed that the equilibrium position peaked in the $Re$ range between 30 and 40 for $d/H \leq$ 0.3,
while the capsule migrated to the channel centerline regardless of $Re$ for $d/H$ = 0.4 (i.e., small channel)~\citep{Shin2011}.
On the other hand,
in a numerical analysis using a three-dimensional spherical capsule model,
\cite{Kilimnik2011} showed that the equilibrium peak position in a rectangular microchannel with $d/H$ = 0.2 tended to increase with channel $Re$ in the range between 1 and 100.
\cite{Schaaf2017} also performed numerical simulations of spherical capsules in a square channel for 0.1 $\leq d/H \leq$ 0.4 and 5 $\leq Re \leq$ 100 without viscosity contrast (i.e., $\lambda$ = 1),
and showed that the equilibrium position was nearly independent of $Re$~\citep{Schaaf2017}.
In a more recent numerical analysis by~\cite{Alghalibi2019},
simulations of a spherical hyperelastic particle in a circular channel with $d/D$ = 0.2 were performed with 100 $\leq Re \leq$ 400 and Weber number 0.125 $ \leq We \leq$ 4.0,
the latter of which is the ratio of the inertial effect to the elastic effect acting on the particles.
Their numerical results showed that regardless of $Re$,
the final equilibrium position of a deformable particle was the centerline,
and harder particles (i.e., with lower $We$) tended to rapidly migrate toward the channel center~\citep{Alghalibi2019}.

Although the equilibrium positions of nonspherical rigid particles have been investigated both by experimental observations~\citep{Masaeli2012} and a numerical simulation~\citep{Huang2017},
the inertial migration of biconcave capsules that model RBCs has not been fully described yet.
Numerical analyses have investigated the behaviour of RBCs under small $Re$ in small microchannels,
with simulations were performed for a viscosity ratio $\lambda$ = 1 and for circular microchannels with 0.3 $< d/D <$ 0.8~\citep{Fedosov2014},
and for a physiologically relevant viscosity ratio $\lambda$ = 5 and a rectangular microchannel with $d/H$ = 0.8~\citep{Guckenberger2018}. 
Despite these efforts,
there has been no comprehensive analysis of the inertial migration of RBCs in large microchannels with diameters of several dozen micrometers $d/D \sim$ 0.1.

Aiming for the precise description of the inertial migration of RBCs in a microchannel, 
we thus performed numerical simulations for individual RBCs with a major diameter of $d$ = 8 $\mu$m,
subject to various $Ca$ in a circular microchannel with $D$ = 50 $\mu$m (i.e., $d/D$ = 0.16).
Each RBC is modeled as a biconcave capsule,
whose membrane follows the Skalak constitutive (SK) law~\citep{Skalak1973}.
Since this problem requires heavy computational resources,
we resort to GPU computing,
using the lattice-Boltzmann method (LBM) for the inner and outer fluids and the finite element method (FEM) to analyse the deformation of the RBC membrane.
This model has been successfully applied to the analysis of multi-RBC interactions in circular microchannels~\citep{Takeishi2014, Takeishi2015, Takeishi2017, Takeishi2019JFM}.
The remainder of this paper is organised as follows.
Section 2 gives the problem statement and numerical methods.
Section 3 presents the numerical results for single RBCs,
and Section 4 presents a discussion,
followed by a summary of the main conclusions in Section 5.
A precise description of membrane mechanics and the numerical setup is presented in the Appendix.

\section{Problem statement}
\subsection{Flow and cell models}
We consider a cellular flow consisting of an external fluid (plasma),
internal fluid (cytoplasm),
and RBC with major diameter $d_0$ (= 2$a_0$ = 8 $\mu$m) and maximum thickness 2 $\mu$m (= $a_0$/2) in a circular channel of diameter with $D$ (= 2$R$ = 50 $\mu$m),
with a resolution of 16 fluid lattices per major radius of RBC (= $a_0$).
The channel length is set to be 20$a_0$, following previous numerical studies~\citep{Fedosov2014, Takeishi2021}.
To show that the channel length is adequate for investigating the behaviour of an RBC that is subject to inertial flow,
we preliminarily assessed the effect of this length on the lateral movement of an RBC in Appendix~\ref{appA3}.
The RBC is modeled as a biconcave capsule,
or a Newtonian fluid enclosed by a thin elastic membrane.

The membrane is modeled as an isotropic and hyperelastic material following the SK law~\citep{Skalak1973}.
The strain energy $w$ of the SK law is given by
\begin{equation}
  w = \frac{G_s}{4} \left( I_1^2 + 2I_1 - 2I_2 + C I_2^2\right),
  \label{SK}
\end{equation}
where $G_s$ is the surface shear elastic modulus,
$C$ is a coefficient representing the area incompressibility,
$I_1 (= \lambda_1^2 + \lambda_2^2 - 2)$ and $I_2 (= \lambda_1^2 \lambda_2^2 -1 = J_s^2 - 1)$ are the first and second invariants of the Green-Lagrange strain tensor,
$\lambda_i$ ($i$ = 1 and 2) are the two principal in-plane stretch ratios,
and $J_s = \lambda_1 \lambda_2$ is the Jacobian,
which expresses the ratio of the deformed to reference surface areas.
In this study, we set $C = 10^2$~\citep{BarthesBiesel2002}.
Bending resistance is also considered~\citep{Li2005},
with a bending modulus $k_b = 5.0 \times 10^{-19}$ J~\citep{Puig-de-Morales-Marinkovic2007}.
These membrane parameters successfully reproduced the deformation of RBCs in shear flow~\citep{Takeishi2014, Takeishi2019JFM},
and also the thickness of the cell-depleted peripheral layer in circular channels~\citep{Takeishi2014}.
We define the initial shape of RBC as a biconcave shape.

Neglecting inertial effects on the membrane deformation,
the static local equilibrium equation of the membrane is given by
\begin{equation}
  \nabla_s \cdot \v{T} + \v{q} = \v{0},
  \label{StrongForm}
\end{equation}
where $\nabla_s (= \left( \v{I} - \v{n} \v{n} \right) \cdot \nabla)$ is the surface gradient operator,
$\v{n}$ is the unit normal outward vector in the deformed state,
$\v{q}$ is the load on the membrane,
and $\v{T}$ is the in-plane elastic tension that is obtained from the SK law (equation~\ref{SK}).
More precise description of membrane mechanics are presented in Appendix~\ref{appA2}. 

It is known that the usual distribution of the hemoglobin concentration in individual RBCs ranges from 27 to 37 g/dl,
corresponding to an internal fluid viscosity of $\mu_1$ = 5--15 cP (= 5--15 mPa$\cdot$s)~\citep{Mohandas2008},
while the normal plasma viscosity is $\mu_0$  = 1.1--1.3 cP (= 1.1--1.3 mPa$\cdot$s) for plasma at 37 $^\circ$C~\citep{Harkness1970}.
Hence, the physiologically relevant viscosity ratio can be taken as $\lambda (= \mu_1/\mu_0)$ = 4.2--12.5 if the plasma viscosity is set to be $\mu_0$ = 1.2 cP.
In our study, therefore, the physiologically relevant viscosity ratio is set to be $\lambda$ = 5. 
Unless otherwise specified,
we show the results obtained with $\lambda$ = 5.

The fluids are modeled as an incompressible Navier--Stokes equation,
with a governing equation of fluid velocity $\v{v}$:
\begin{align}
	\rho \left( \frac{\partial \v{v}}{\partial t} + \v{v} \cdot \nabla \v{v} \right)
	&= \nabla \cdot \v{\sigma}^f  + \rho \v{f}, \\
	\nabla \cdot \v{v} &= 0,
\end{align}
and 
\begin{align}
\v{\sigma}^f = -p\v{I} + \mu \left( \nabla \v{v} + \nabla \v{v}^T \right),
\end{align}
where $\v{\sigma}^f$ is the total stress tensor of the flow,
$p$ is the pressure, 
$\rho$ is the fluid density,
$\v{f}$ is the body force,
and $\mu$ is the viscosity of liquids,
which is expressed using a volume fraction of the inner fluid $\alpha$ (0 $\leq \alpha \leq$ 1) as:
\begin{align}
        \mu = \left\{ 1 + \left( \lambda - 1 \right) \alpha \right\} \mu_0.
\end{align}
The dynamic condition requires that the load $\v{q}$ to be equal to the traction jump $\left( \v{\sigma}^f_{out} - \v{\sigma}^f_{in} \right)$ across the membrane:
\begin{align}
	\v{q} = \left( \v{\sigma}^f_\mathrm{out} - \v{\sigma}^f_\mathrm{in} \right) \cdot \v{n},
\end{align}
where the subscripts `out' and `in', respectively, represent the outer and internal regions of the capsule,
and $\v{n}$ is the unit normal outward vector in the deformed state.

The problem is characterised by Reynolds number $Re$ and capillary number $Ca$:
\begin{align}
  &Re = \frac{\rho D V_{\mathrm{max}}^{\infty}}{\mu_0}, \\
  &Ca = \frac{\mu_0 \dot{\gamma}_\mathrm{m} a_0}{G_s} = \frac{\mu_0 V_{\mathrm{max}}^{\infty}}{G_s} \frac{a_0}{4 R},
\end{align}
where $V_{\mathrm{max}}^{\infty} (= 2V_\mathrm{m}^{\infty})$ is the maximum plasma velocity in the absence of any cells,
and $\dot{\gamma}_\mathrm{m} (= V_\mathrm{m}^{\infty}/D)$ is the mean shear rate.
Increasing $Re$ under constant $Ca$ corresponds to increasing $G_s$, namely, a harder RBC.

\subsection{Numerical simulation and setup}
The governing equations for the fluid are discretised by the LBM based on the D3Q19 model~\citep{Chen1998}.
We track the Lagrangian points of the membrane material points $\v{x}(\v{X},t)$ over time,
where $\v{X}$ is a material point on the membrane in the reference state.
Based on the virtual work principle,
the above strong-form equation (\ref{StrongForm}) can be rewritten in weak form as 
\begin{equation}
  \int_S \v{\hat{u}} \cdot \v{q} dS = \int_S \v{\hat{\epsilon}} : \v{T} dS,
  \label{WeakForm}
\end{equation}
where $\v{\hat{u}}$ and $\v{\hat{\epsilon}} = ( \nabla_s \v{\hat{u}} + \nabla_s \v{\hat{u}}^T )\big/2$ are the virtual displacement and virtual strain, respectively.
The FEM is used to solve equation (\ref{WeakForm}) and obtain the load $\v{q}$ acting on the membrane~\citep{Walter2010}.
The velocity at the membrane node is obtained by interpolating the velocities at the fluid node using the immersed boundary method~\citep{Peskin2002}.
The membrane node is updated by Lagrangian tracking with the no-slip condition.
The explicit fourth-order Runge--Kutta method is used for the time integration. 
The volume-of-fluid method~\citep{Yokoi2007} and front-tracking method~\citep{Unverdi1992} are employed to update the viscosity in the fluid lattices.
A volume constraint is implemented to counteract the accumulation of small errors in the volume of the individual cells~\citep{Freund2007}:
in our simulation, the volume error is always maintained lower than $1.0 \times 10^{-3}$\%,
as tested and validated in our previous study of cell flow in circular channels~\citep{Takeishi2016}.
All procedures were fully implemented on a GPU to accelerate the numerical simulation.
More precise explanations are provided in our previous work~\citep[see also][]{Takeishi2019JFM}.

The channel flow is generated by a pressure gradient.
Periodic boundary conditions are imposed on flow direction ($z$-direction).
No-slip conditions are employed for the walls (radial direction).
The resolution that we set has been shown to successfully represent single- and multi-cellular dynamics~\citep{Takeishi2014, Takeishi2019JFM, Takeishi2021};
the mesh size of the LBM for the fluid solution was set to be 250 nm,
and that of the finite elements describing the membrane was approximately 250 nm (an unstructured mesh with 5,120 elements was used for the FEM).
This resolution has been shown to successfully represent single- and multi-cellular dynamics~\citep{Takeishi2014};
also, the results of multi-cellular dynamics are not changed by using twice the resolution for both the fluid and membrane meshes~\citep[see also][]{Takeishi2014}.

\subsection{Analysis}
To quantify the effects of the radial position of the RBC centroid and the shape of the deformed cell on fluid flow,
the power (or energy dissipation) associated with membrane deformations is considered,
and is given by
\begin{align}
  \delta W_{\mathrm{mem}}
  &= \int \hat{\v{q}} \cdot \left( \v{v}^{(m)} - \v{V}^\infty (r) \right) dS, \\
  \delta W_\mathrm{mem}
  &= \frac{\mu_0 D V_\mathrm{max}^{\infty 2}}{2} \int \hat{\v{q}}^\ast \cdot \left( \v{v}^{(m)\ast} - \v{V}^{\infty \ast} (r) \right) dS^\ast, \nonumber \\
  &= \frac{\mu_0 D V_\mathrm{max}^{\infty 2}}{2} \int \left[ \hat{q}_x^\ast v_x^{(m) \ast} + \hat{q}_y^\ast v_y^{(m) \ast} + \hat{q}_z^\ast \left( v_z^{(m) \ast} - {V}_z^{\infty \ast} (r) \right) \right] dS^\ast, \\
  \to \delta W_\mathrm{mem}^\ast
  &= \delta W_\mathrm{mem}/\left( \mu_0 D V_\mathrm{max}^{\infty 2}/2 \right),
  \label{ediss_mem}
\end{align}
where $\v{V}^\infty (r) = \left(0, 0, V_\mathrm{max}^\infty \left[ 1 - (r/R)^2\right] \right)$ is the fluid flow velocity without cells,
$\hat{\v{q}}$ is the load acting on the membrane and includes the contribution of bending rigidity,
$r$ is the membrane distance from the channel center, 
$\v{v}^{(m)}$ is the interfacial velocity of the membrane,
and $S$ is the membrane surface area.
Here, non-dimensional variables are defined as $\hat{\v{q}}^\ast = \hat{\v{q}}/(\mu_0 \dot{\gamma}_\mathrm{m})$,
$\v{v}^{(m) \ast} = \v{v}^{(m)}/V_{\mathrm{max}}^\infty$,
$\v{V}^{\infty \ast} = \v{V}^{\infty}/V_{\mathrm{max}}^\infty$, and
$S^\ast = S/D^2$.
\begin{figure}
  \centering
  \includegraphics[height=4.5cm]{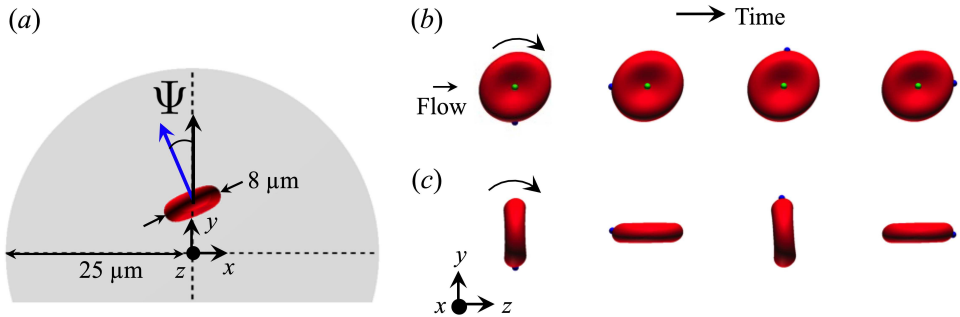}
  \caption{
	  ($a$) Representative snapshot of the RBC orientation angle with $\Psi$ on the cross-sectional area of the channel,
	  where $\Psi$ is the angle between the radial direction toward the RBC centroid and the normal vector at the initial concave node point.
	  ($b$ and $c$) Snapshots of representative ($b$) stable rolling motion with $|\Psi_\infty| \sim \pi$/2,
	  and ($c$) tumbling motion with $|\Psi_\infty| \sim$ 0 and $\pi$,
	  where green dots represent material points at the initial concave node point,
	  and blue dots represent those at the initial edge node point.
	  Flow direction is from {\it left} to {\it right}.
  }
  \label{fig:ori_angle}
\end{figure}

For the following analysis,
the behaviour of RBCs in the channel is quantified by an orientation angle $\Psi$ on the cross-sectional area of the channel as shown in figure~\ref{fig:ori_angle}($a$),
where $\Psi$ is the angle between the radial direction toward the RBC centroid and the normal vector at the initial concave node point.
Following previous numerical studies~\citep{Takeishi2019JFM, Takeishi2021},
we define two types of RBC flow modes depending on equilibrium orientation angle $\Psi_\infty$.
If $\Psi$ orients perpendicular to the radial direction, i.e., $|\Psi_\infty| \sim \pi$/2,
showing a wheel-like configuration,
the flow mode is defined as a rolling motion (see figure~\ref{fig:ori_angle}$b$).
On the other hand,
if $\Psi$ orients parallel to the radial direction, i.e., $|\Psi_\infty| \sim$ 0 and $\pi$,
showing a flipping motion with cyclic,
the flow mode is defined as a tumbling motion (see figure~\ref{fig:ori_angle}$c$).
More detailed transitions of $\Psi$ in each mode are described in below (see figure~\ref{fig:map_ori}).
Time averaging starts after the orientation angle and radial position of RBC reach their final values,
and the time averaging size is usually set to be $\dot{\gamma}_\mathrm{m} t \geq$ 10$^2$.

\section{Results}
\subsection{Effect of initial orientation angle $\Psi_0$ on stable flow mode}
\begin{figure}
  \centering
  \includegraphics[height=5.5cm]{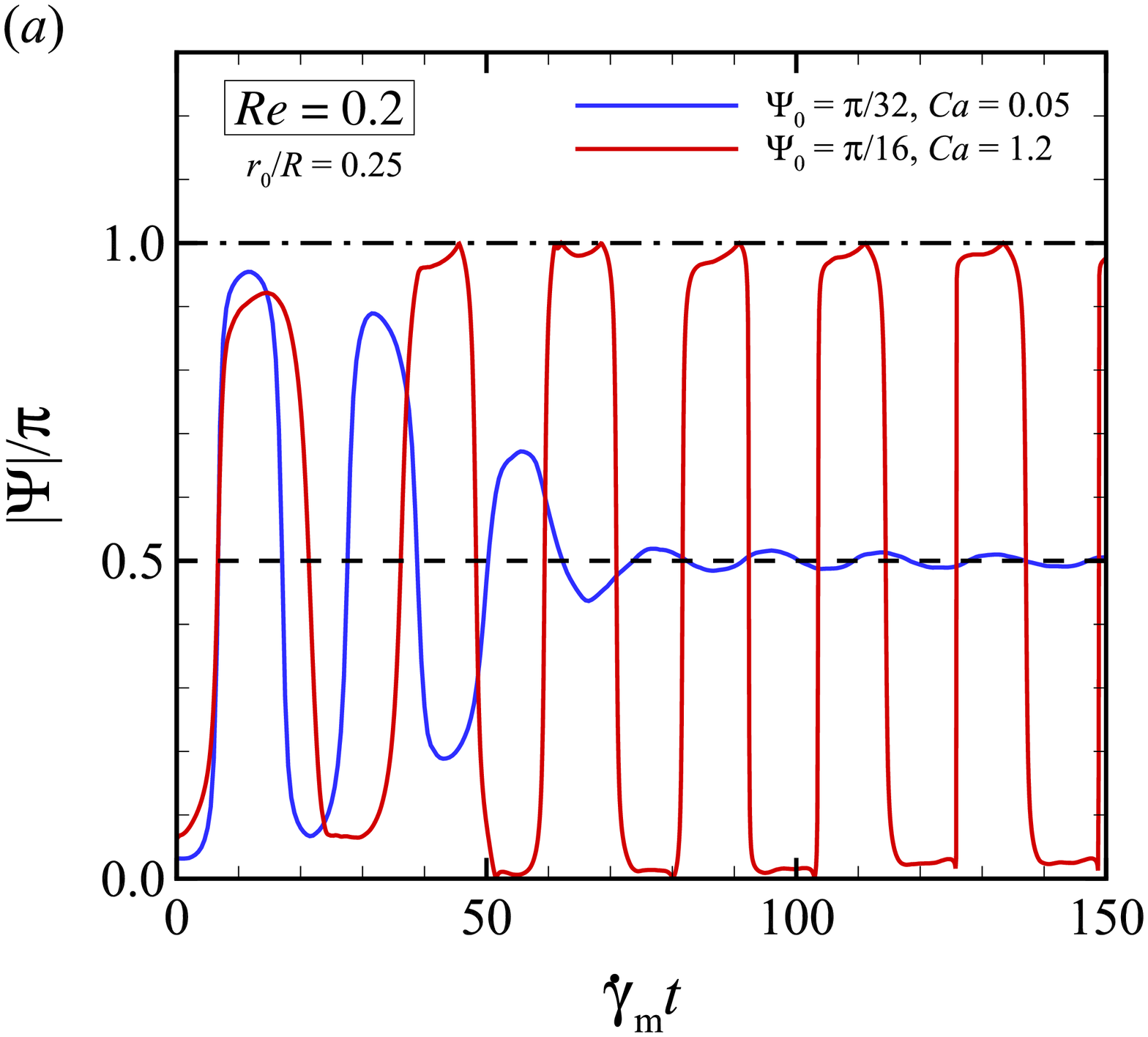}
  \includegraphics[height=5.5cm]{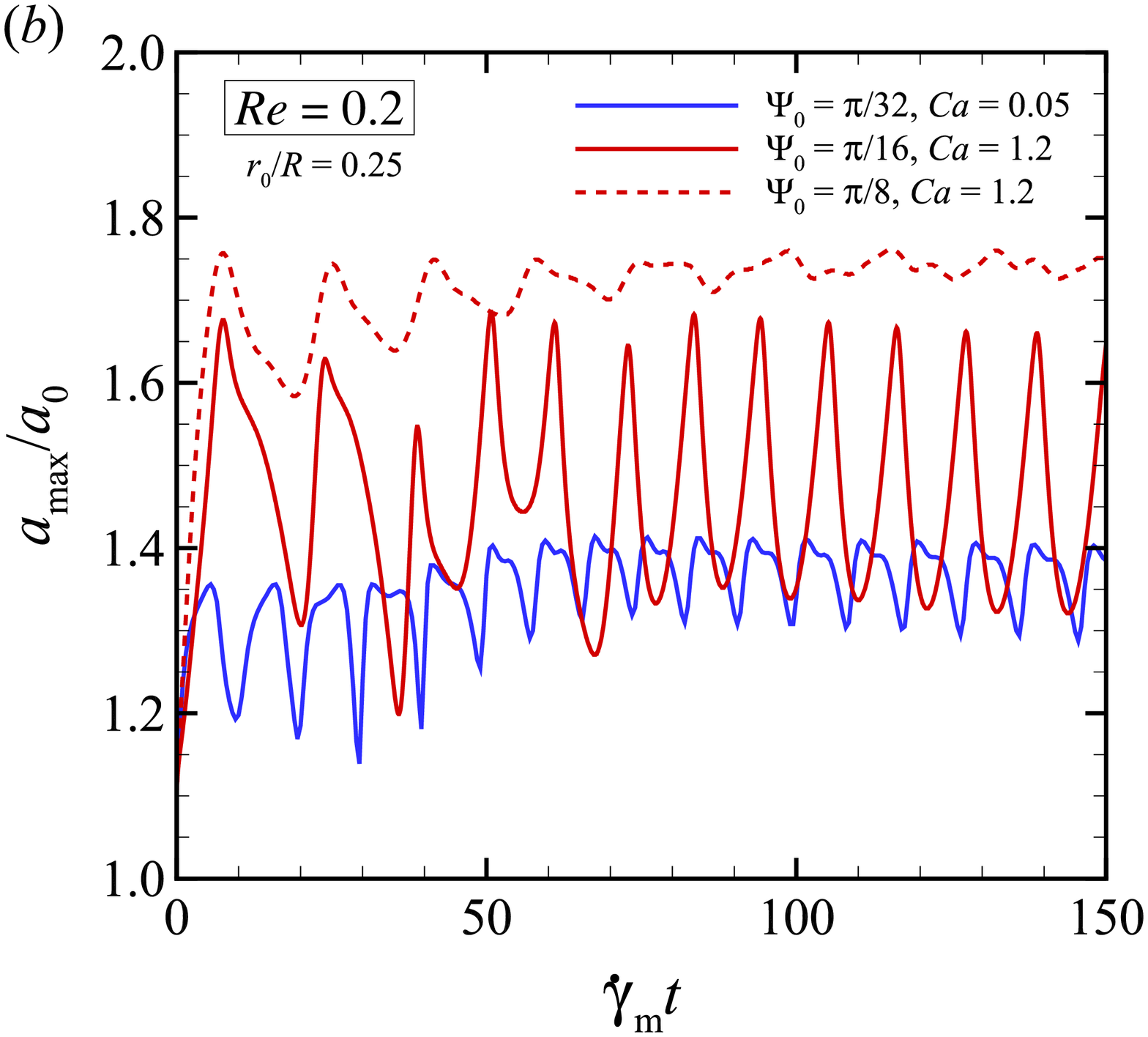}  
  \includegraphics[height=5.5cm]{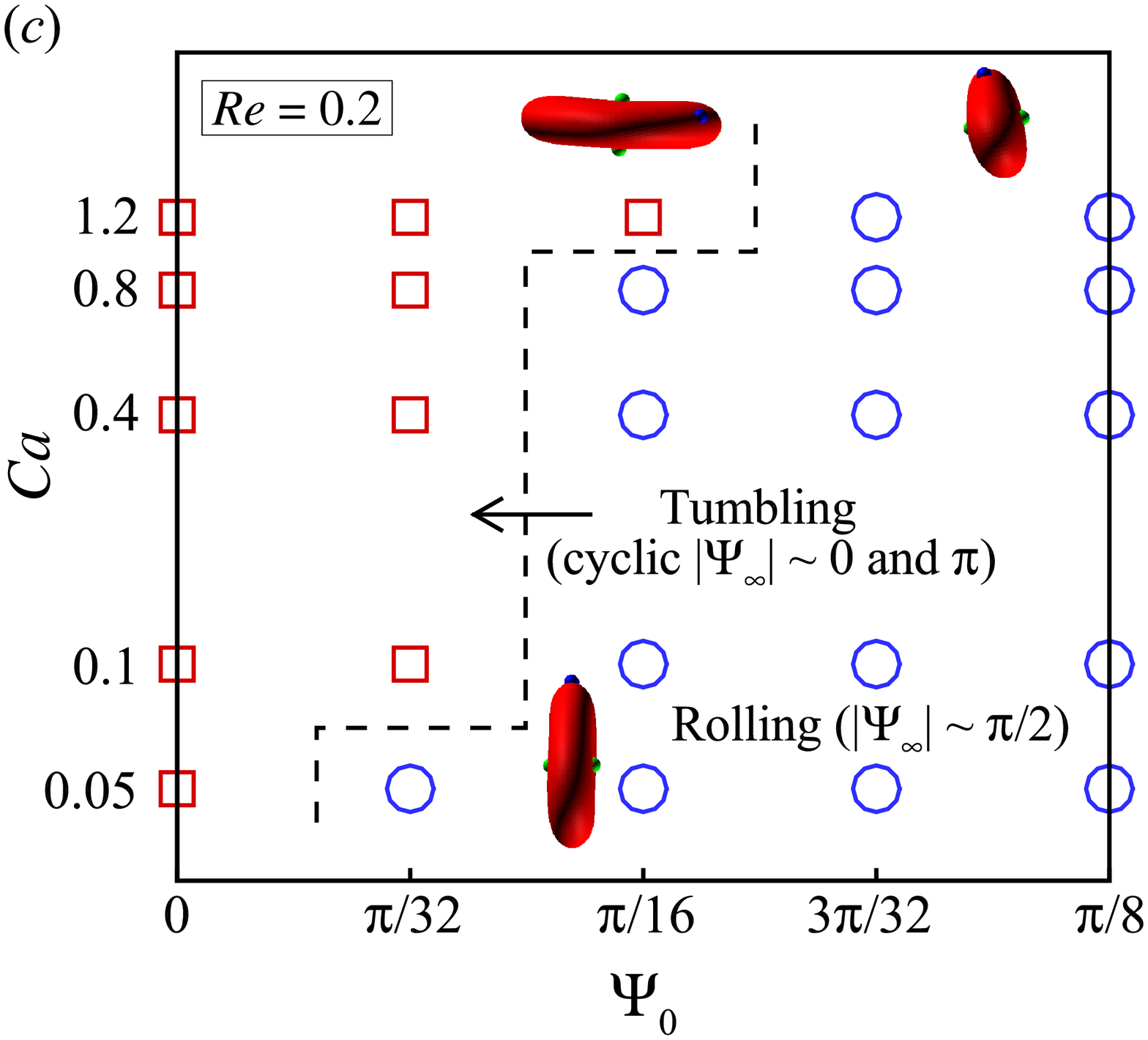}
  \includegraphics[height=5.5cm]{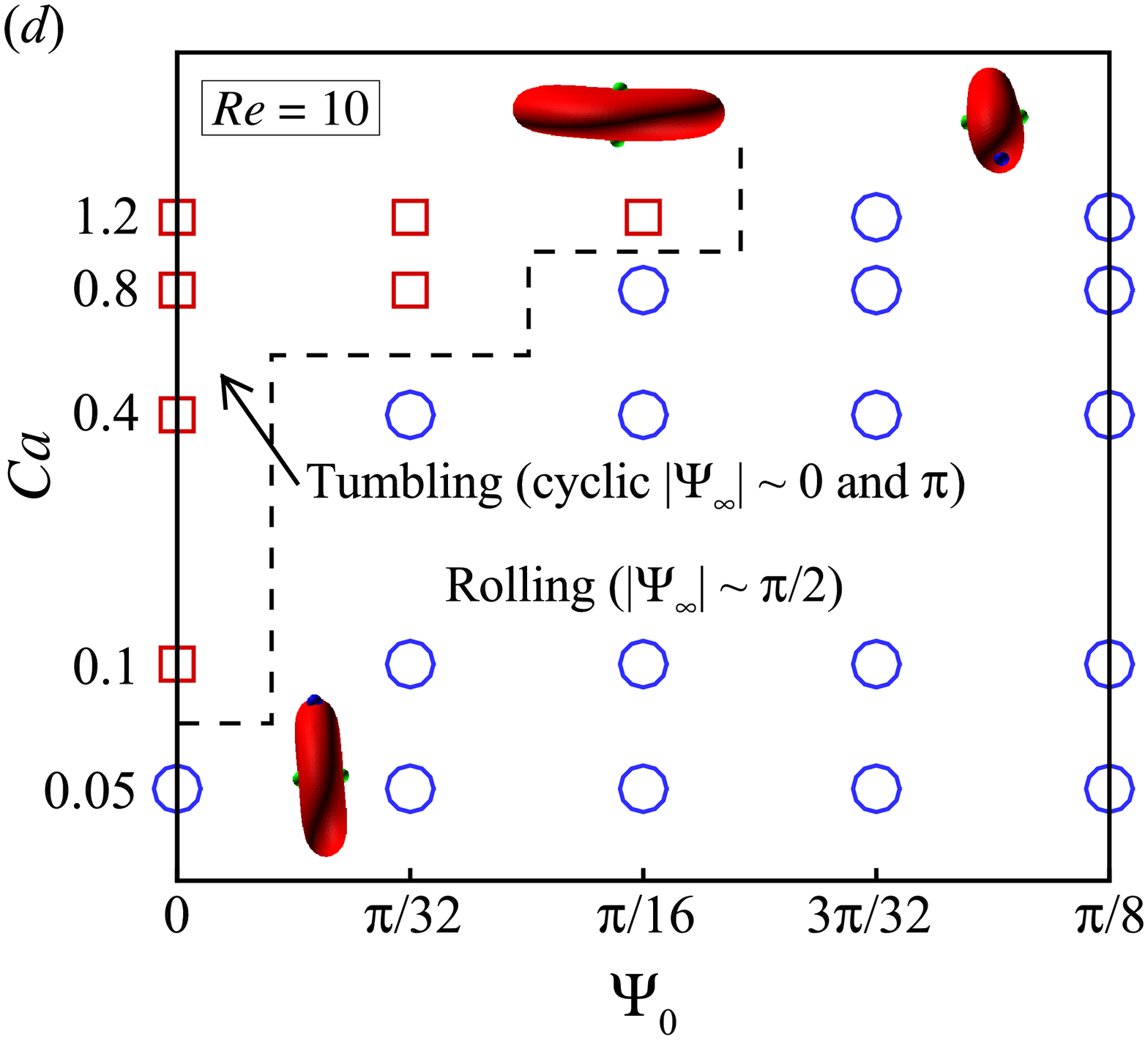}
  \caption{
         ($a$) Time history of the orientation angle $|\Psi|/\pi$ on the cross-sectional ($x$-$y$) plane for different initial orientation angles $\Psi_0$ (= $\pi$/32 and $\pi$/16) and different $Ca$ (= 0.05 and 1.2),
         where the dashed line and dash-dot line represent $|\Psi|$ = $\pi$/2 and $|\Psi|$ = $\pi$, respectively
         (see the supplementary movies 1 and 2, at \textit{https://doi.org/xxx.yyy.zzz}).
         ($b$) Time history of the deformation index $a_\mathrm{max}/a_0$ for different $\Psi_0$ (= $\pi$/32, $\pi$/16, and $\pi$/8) and different $Ca$ (= 0.05 and 1.2).
          The results are obtained with low $Re$ (= 0.2).
	  ($c$ and $d$) Diagram of the final stable orientation of the RBC subject to different $Ca$ for different initial orientations $\Psi_0$.
	  The results for $Re$ = 0.2 and 10 are reported in ($c$) and ($d$), respectively.
	  The insets represent the steady state of the RBC at the equilibrium orientation $|\Psi_\infty|$.
	  Circles denote stable rolling motion ($|\Psi_\infty| \sim \pi/2$),
	  and squares denote stable tumbling motion (cyclic $|\Psi_\infty| \sim$ 0 and $\pi$).
 	  All results are obtained with $r_0/R$ = 0.25 and $\lambda$ = 5.
  }
  \label{fig:map_ori}
\end{figure}

We first investigate the equilibrium orientation angle $\Psi_\infty$ on the cross-sectional ($x$-$y$) plane depending on the initial orientation angle $\Psi_0$.
Simulations are started from a slightly off-centered radial position,
with the radial position of the RBC centroid set as $r_0/R$ = 0.25,
where $r_0$ is the distance from the channel center to the RBC centroid on the cross-sectional plane.
The time history of $\Psi$ for different initial orientation angles $\Psi_0$ (= $\pi$/32 and $\pi$/16) and different $Ca$ (= 0.05 and 1.2) are shown in figure~\ref{fig:map_ori}($a$).
The results are obtained with low $Re$ (= 0.2),
and can be assumed to be almost inertialess~\citep{Takeishi2019JFM, Takeishi2021}.
An RBC that is subject to low $Ca$ (= 0.05) and that is only slightly tilted, with $\Psi_0$ = $\pi$/32,
gradually orients parallel to the radial plane,
showing a wheel-like configuration,
the so-called rolling motion with $|\Psi_\infty| \sim \pi$/2 (blue line in figure~\ref{fig:map_ori}$a$) (see the supplementary movie 1).
In contrast, an RBC that is subject to high $Ca$ (= 1.2) and that is further tilted, with $\Psi_0$ = $\pi$/16, immediately shows a flipping motion,
with with cyclic $|\Psi_\infty| \sim$ 0 and $\pi$,
the so-called tumbling motion (red line in figure~\ref{fig:map_ori}$a$) (see the supplementary movie 2).
The characteristic time histories in these modes persist even when $Re$ increases from 0.2 to 10 (data is not shown).
The results further show that the aforementioned two different types of modes have completed at relatively early time periods $O(\dot{\gamma}_\mathrm{m} t) \leq 10^2$ (see also figure~\ref{fig:timehist_re10_ca}$b$).
Oriented RBCs, however, are still migrating towards the radial direction in each stable mode.

The degree of cell deformation is quantified by the maximal radius $a_\mathrm{max}$ of deformed RBCs,
and is obtained from the eigenvalues of the inertia tensor of an equivalent ellipsoid approximating deformed RBCs~\citep{Ramanujan1998}.
The time history of $a_\mathrm{max}/a_0$ is shown in figure~\ref{fig:map_ori}($b$).
The tumbling RBC exhibits large cyclic extension with the same period as its rotations (figure~\ref{fig:map_ori}$b$, red solid line),
while the rolling RBC has relatively small fluctuations in $a_\mathrm{max}/a_0$ (figure~\ref{fig:map_ori}$b$, blue solid line).
For the same $Ca$ (= 1.2),
the rolling RBC exhibits greater extension than the tumbling RBC (figure~\ref{fig:map_ori}$b$, red dashed line).

The stable orientation $\Psi_\infty$ of RBCs is investigated for different initial orientation angles $\Psi_0$, different $Ca$ (= 0.05--1.2), and different $Re$ (= 0.2 and 10),
and the results are summarized in figures~\ref{fig:map_ori}($c$) and \ref{fig:map_ori}($d$).
At low $Re$ (= 0.2),
RBCs subject to low $Ca$ (= 0.05) have a rolling motion ($|\Psi_\infty| \sim \pi$/2) for $\Psi_0 \geq \pi$/32 (figure~\ref{fig:map_ori}$c$).
This result suggests that most RBCs tend to have a rolling motion.
As $Ca$ increases,
RBCs tend to have a tumbling motion,
for $\Psi_0 \leq \pi$/16 (figure~\ref{fig:map_ori}$c$).
Since the rolling motion of RBCs subject to $Ca$ = 1.2 is observed at least for $\Psi_0 \geq 3\pi/32$, 
the dominant flow mode is still a rolling motion even at high $Ca$ (figure~\ref{fig:map_ori}$c$).

At high $Re$ (=10),
all RBCs have a rolling motion even at low $Ca$ (= 0.05),
as shown in figure~\ref{fig:map_ori}($d$).
Hence, finite inertia imposes disturbances on the membrane,
which potentially allows RBCs to have a stable rolling motion.
For higher $Ca$ (= 1.2),
the final stable mode depends on initial orientation angle $\Psi_0$ but remains the same as in the case with almost no inertia ($Re$ = 0.2),
where the tumbling motion appears for $0 \leq \Psi_0 \leq \pi/16$ and the rolling motion is seen for $\Psi_0 \geq 3\pi/32$ (figure~\ref{fig:map_ori}$d$).
Comparing with the case of low $Re$ = 0.2,
the higher $Re$ conditions impede the tumbling motion (figures~\ref{fig:map_ori}$c$ and \ref{fig:map_ori}$d$).

\subsection{Effect of capillary number $Ca$ on equilibrium radial position}
\begin{figure}
  \centering
  \includegraphics[height=4cm]{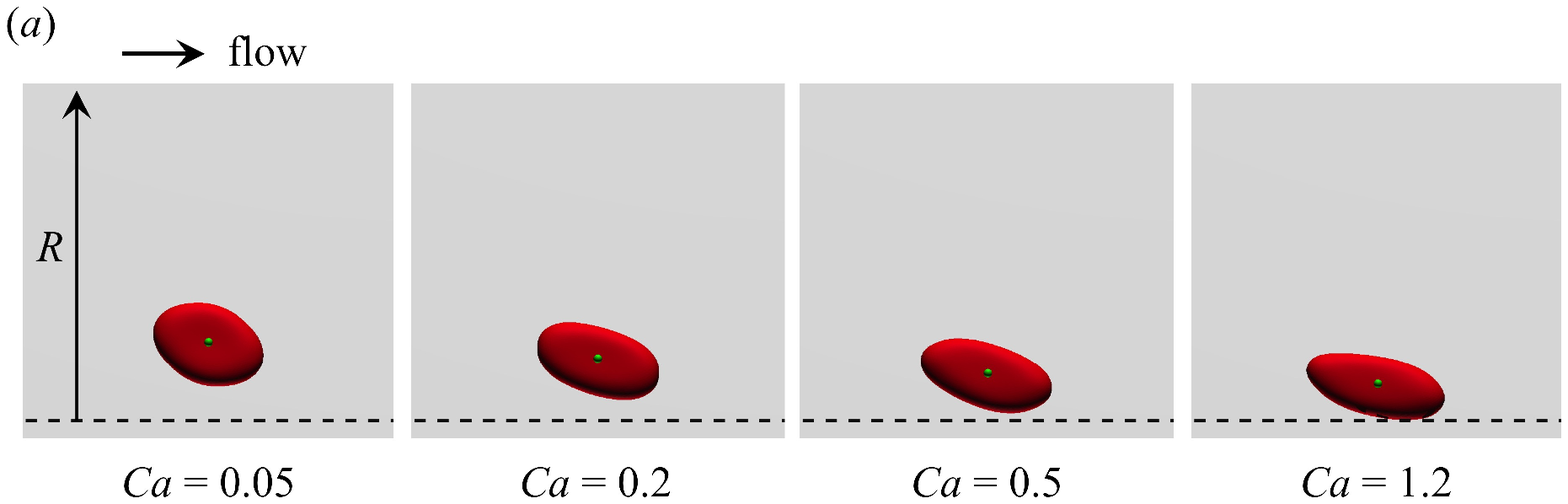}
  \includegraphics[height=5.5cm]{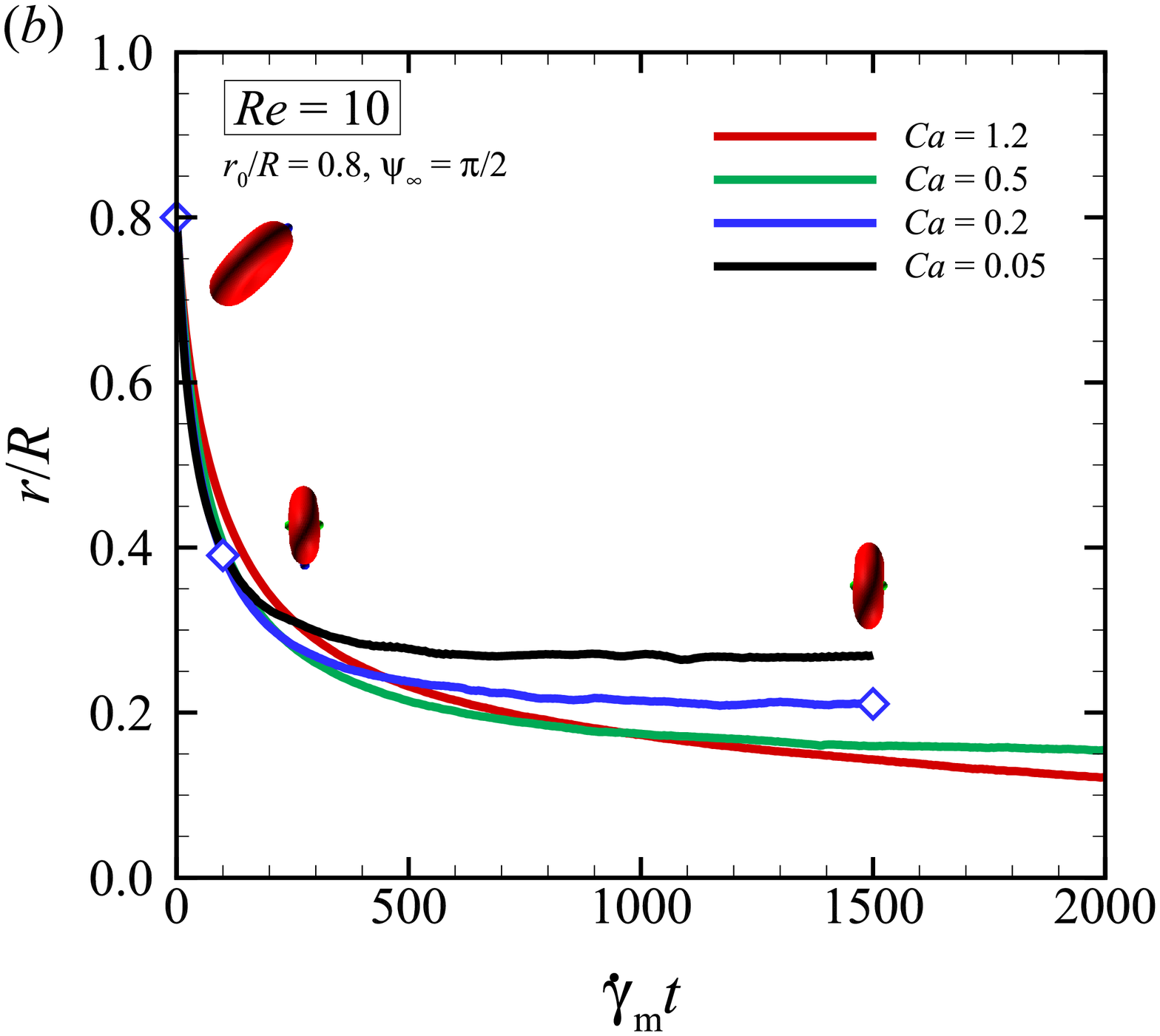}
  \includegraphics[height=5.5cm]{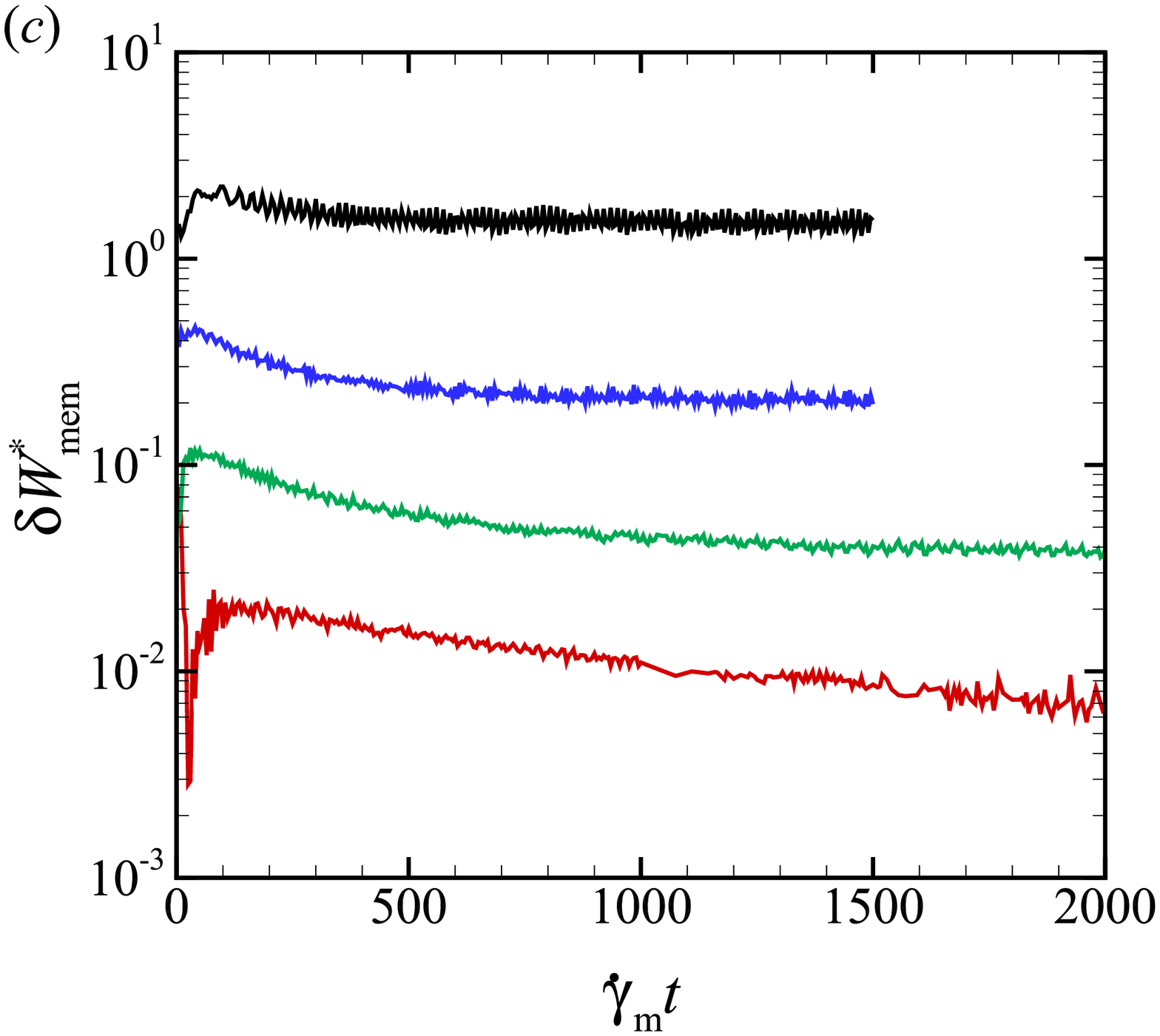}
  \caption{
	  ($a$) Snapshots of a rolling RBC ($|\Psi_\infty| \sim \pi$/2) subject to different $Ca$ (= 0.05, 0.2, 0.5, and 1.2) in $Re$ = 10,
	  where results are at $\dot{\gamma}_\mathrm{m}t$ = 1500 for $Ca$ = 0.05 and 0.2,
	  and at $\dot{\gamma}_\mathrm{m} t$ = 2000 for $Ca$ = 0.5 and 1.2.
	  The RBCs start from the near-wall position $r_0/R$ = 0.8.
         The temporal changes corresponding to these snapshots are shown in the supplementary movies:
         movie 3 for $Ca$ = 0.05,
         movie 4 for $Ca$ = 0.2,
         and movie 5 for $Ca$ = 1.2, respectively.
	  ($b$) Time history of the radial position of the RBC centroid $r/R$,
	  where insets represent snapshots of the axial view of an RBC subject to $Ca$ = 0.2 (at $\dot{\gamma}_\mathrm{m} t$ = 0, 100, and 1500).
	  ($c$) Time history of the powers associated with membrane deformation $\delta W_\mathrm{mem}^\ast$.
  }
  \label{fig:timehist_re10_ca}
\end{figure}

Next, we investigate the equilibrium radial position of an RBC centroid at $Re$ = 10 for different $Ca$.
Figure~\ref{fig:timehist_re10_ca}($a$) shows snapshots of stable rolling RBCs ($|\Psi_\infty| \sim \pi/2$),
each subject to different $Ca$,
when they have reached to each equilibrium radial position (figure~\ref{fig:timehist_re10_ca}$b$).
All RBCs starts from the near-wall position $r_0/R$ = 0.8 with the initial orientation angle $\Psi_0$ = $\pi$/4.
As $Ca$ increases,
the RBCs are extended to the flow direction (figure~\ref{fig:timehist_re10_ca}$a$).
Figure~\ref{fig:timehist_re10_ca}($b$) is the time history of the RBC centroid for different $Ca$,
where insets represent snapshots of the axial view of an RBC subject to low $Ca$ (= 0.2) at specific time points $\dot{\gamma}_\mathrm{m} t$ = 0, 100, and 1500 (see the supplementary movie 4).
At low $Ca$ (= 0.2),
the RBC exhibits a stable rolling motion within a relatively early time period $\dot{\gamma}_\mathrm{m} t \leq$ 100 (see insets in figure~\ref{fig:timehist_re10_ca}$b$),
and then the RBC attains the radial position $r/R \approx$ 0.2.
At the highest $Ca$ (= 1.2) in this paper,
the RBC is still migrating towards the channel center even after time $\dot{\gamma}_\mathrm{m} t$ = 2000 (figure~\ref{fig:timehist_re10_ca}$b$) (see supplementary movie 5).
Note that the volumetric flow rate,
which is inversely proportional to the apparent viscosity,
remains almost the same independent of the equilibrium radial position (data not shown),
i.e., the flow resistances are not significantly changed by RBC deformation.

The time history of powers associated with membrane deformations $\delta W_\mathrm{mem}^\ast$ are shown in figure~\ref{fig:timehist_re10_ca}($c$).
The result shows that powers $\delta W_\mathrm{mem}^\ast$ tend to decrease as $Ca$ increases.
Although $\delta W_\mathrm{mem}^\ast$ is still decreasing at the highest $Ca$ (= 1.2),
the power basically retains the same order of magnitude for each $Ca$ (figure~\ref{fig:timehist_re10_ca}$c$) after the onset of stable rolling motion.

\begin{figure}
  \centering
  \includegraphics[height=5.5cm]{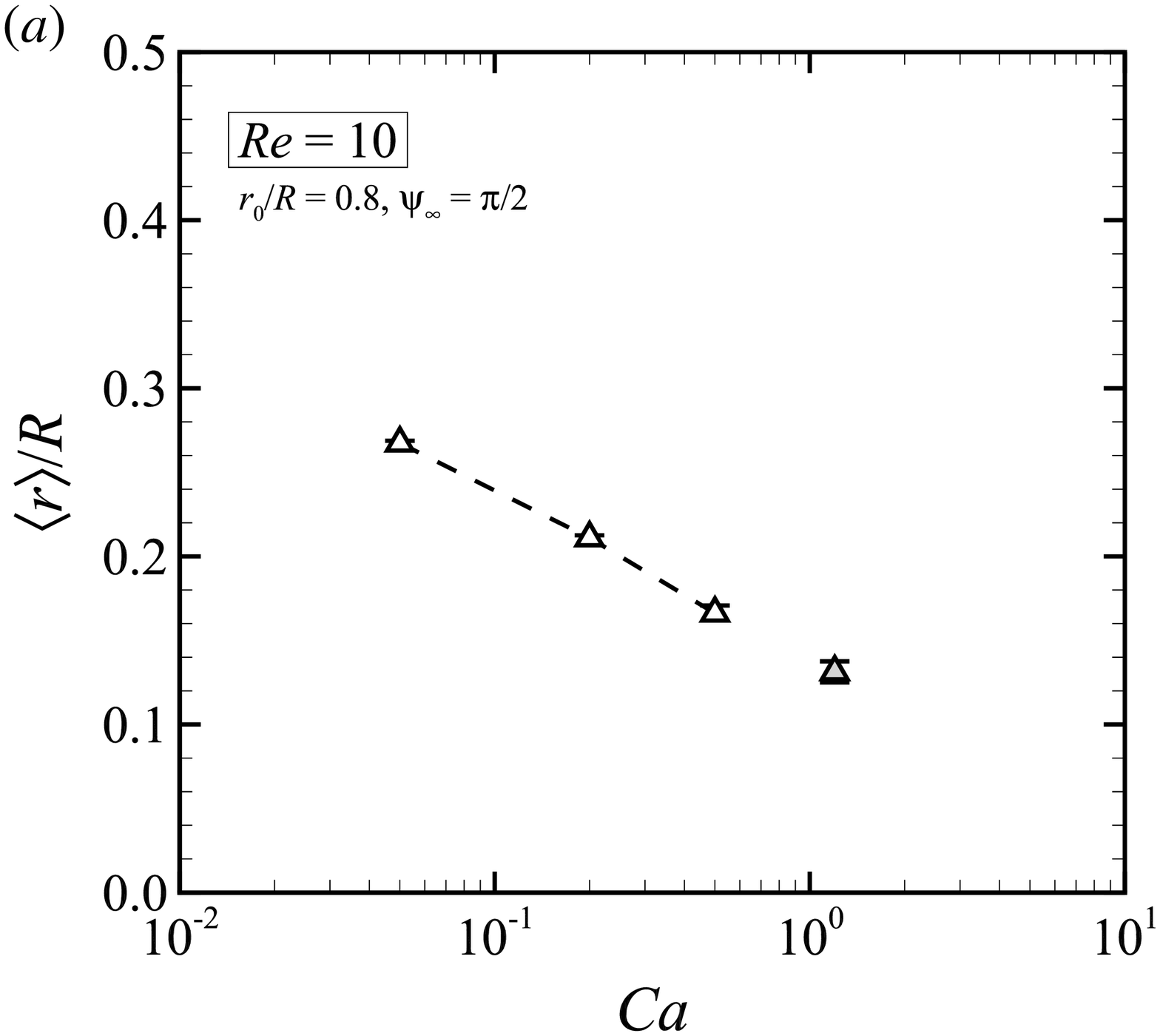}
  \includegraphics[height=5.5cm]{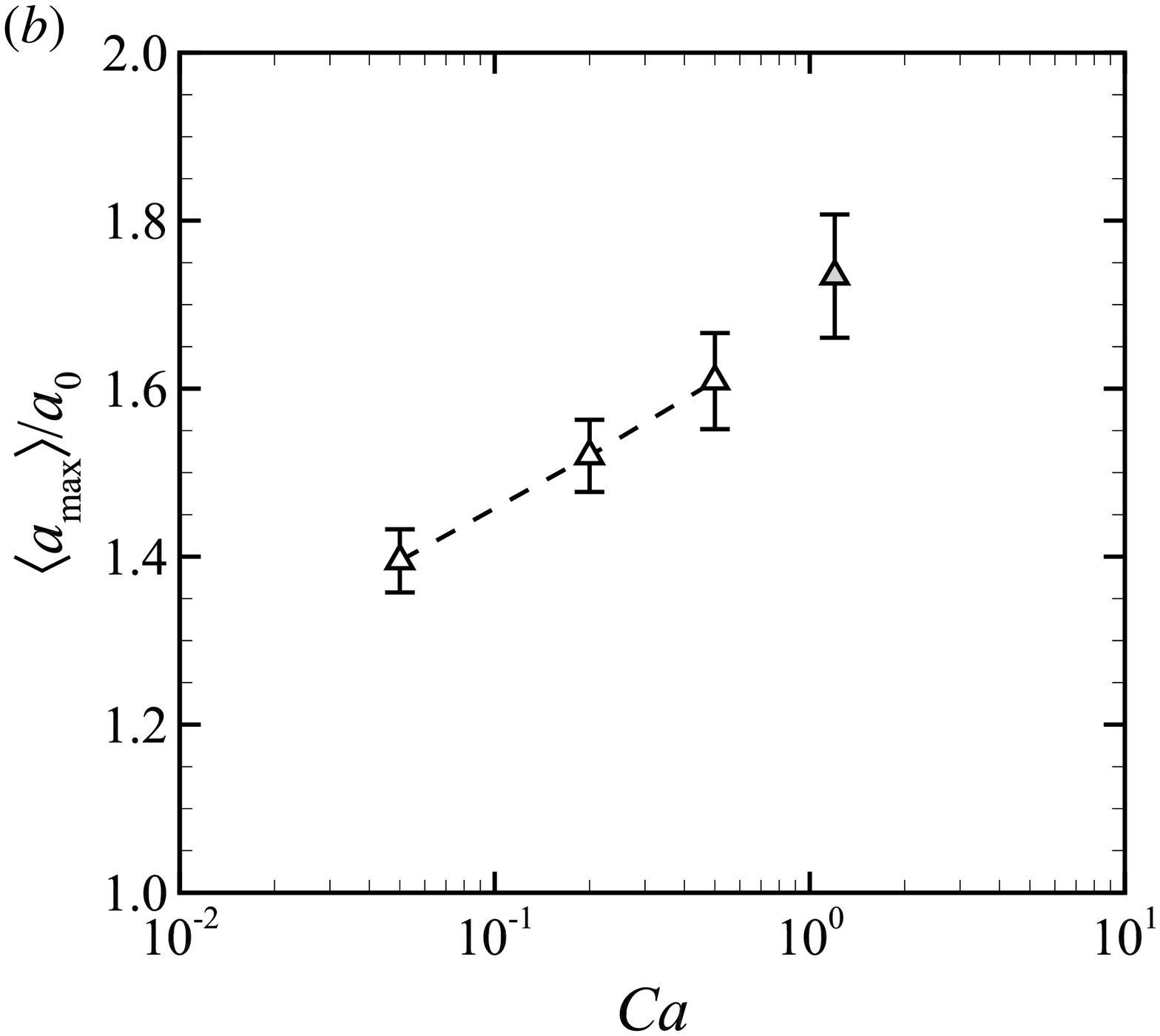}
  \includegraphics[height=5.5cm]{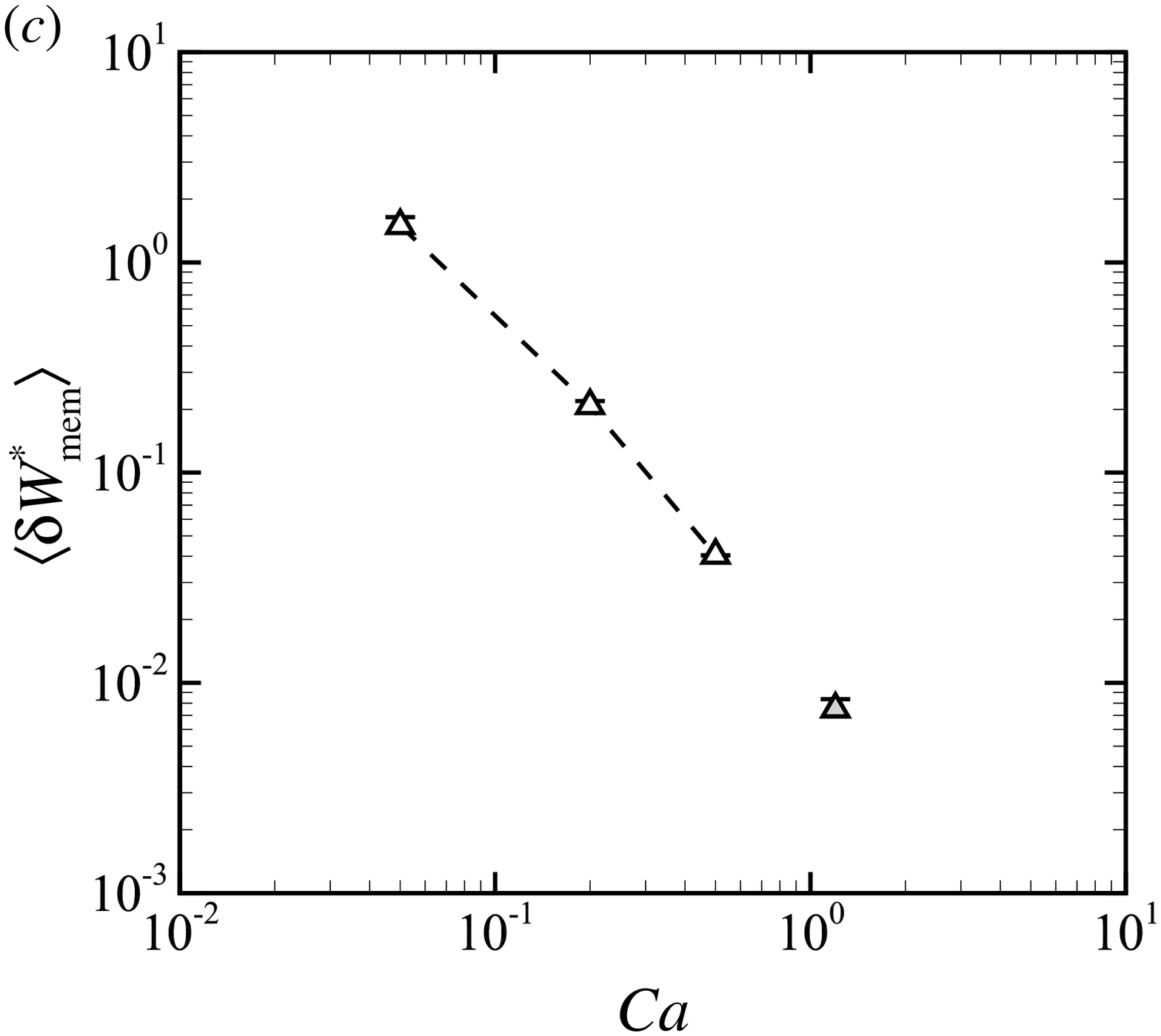}
  \includegraphics[height=5.5cm]{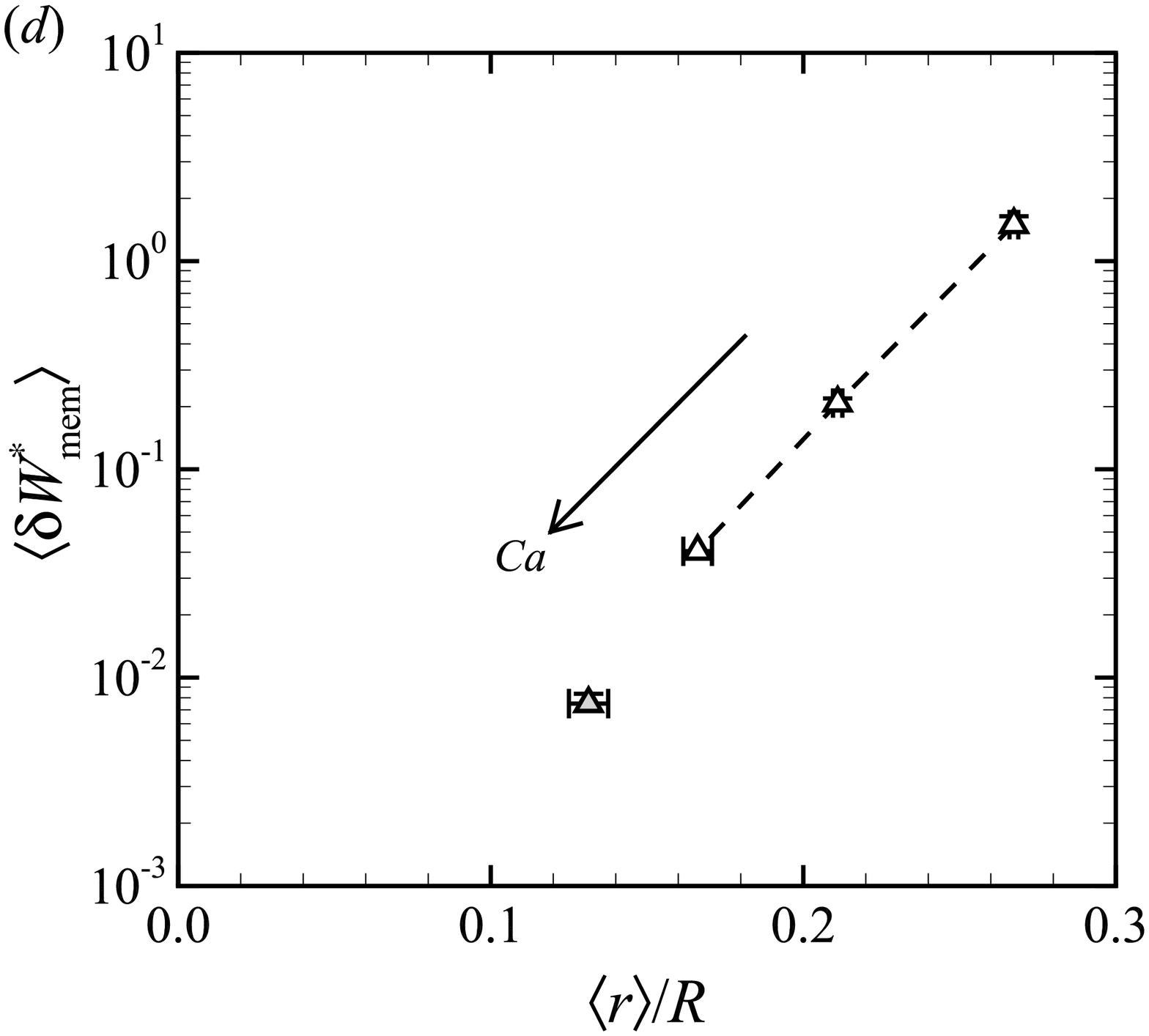}
  \caption{
	  ($a$) Time average of the radial position $\langle r \rangle/R$ of a RBC centroid, 
	  ($b$) the maximum length $\langle a_{\mathrm{max}} \rangle/a_0$ of a deformed RBC, and
	  ($c$) powers $\langle \delta W_{\mathrm{mem}}^\ast \rangle$ associated with membrane deformations occurring with a stable rolling motion ($|\Psi_\infty| \sim \pi$/2).
	  All are presented as a function of $Ca$.
	  ($d$) Replotted data of $\langle \delta W_{\mathrm{mem}}^\ast \rangle$ as a function of equilibrium radial position $\langle r \rangle/R$ for different $Ca$.
	  Results at $Ca$ = 1.2 are for an RBC in transient (still migrating toward the channel center),
	  which is represented by light gray dots.
	  All results are obtained with $Re$ = 10 and $r_0/R$ = 0.8.
  }
  \label{fig:effect_ca}
\end{figure}

Figure~\ref{fig:effect_ca}($a$) shows the time average of the radial position $\langle r \rangle/R$ as a function of $Ca$,
where the error bars represent standard deviations along the time axis (i.e., time fluctuation).
Hereafter, $\langle \cdot \rangle$ denotes the time average.
As $Ca$ increases,
RBCs tend to migrate towards the channel center,
and thus inertial migration is impeded by deformability (figure~\ref{fig:effect_ca}$a$).
Since the RBC subject to the highest $Ca$ (= 1.2) is still in axial migration,
the average radial position $\langle r \rangle/R$ is obtained with data from $\dot{\gamma}_\mathrm{m} t$ = 1500--2000.
Figure~\ref{fig:effect_ca}($b$) indicates the time average of deformation index $\langle a_\mathrm{max} \rangle/a_0$,
and shows that $\langle a_\mathrm{max} \rangle/a_0$ increases with $Ca$ (figure~\ref{fig:effect_ca}$b$).

The powers $\langle \delta W_\mathrm{mem}^\ast \rangle$ as a function of $Ca$ are shown in figure~\ref{fig:effect_ca}($c$).
The powers $\langle \delta W_\mathrm{mem}^\ast \rangle$ decrease as $Ca$ increases,
which is similar in tendency to the radial position $\langle r \rangle/R$ (figure~\ref{fig:effect_ca}$a$) and is opposite to that of the maximum radius $\langle a_\mathrm{max} \rangle/a_0$ (figure~\ref{fig:effect_ca}$b$).
The relationship between $\langle \delta W_\mathrm{mem}^\ast \rangle$ and $\langle r \rangle/R$ is replotted in figure~\ref{fig:effect_ca}($d$).
The order of magnitude of the power $\langle \delta W_\mathrm{mem}^\ast \rangle$ decreases as the equilibrium radial position $\langle r \rangle/R$ decreases (figure~\ref{fig:effect_ca}$d$).

\subsection{Effects of flow modes and initial positions $r_0/R$ on equilibrium radial position}
To clarify how the aforementioned flow modes affect the equilibrium radial positions of RBC centroids,
we investigate the effect of the stable flow modes of RBCs on equilibrium radial positions $\langle r \rangle/R$ as well as the power $\langle \delta W_{\mathrm{mem}}^\ast \rangle$ under specific $Re$ = 10 and $Ca$ = 1.2.
The effects of the initial radial position $r_0$ on $\langle r \rangle/R$ and $\langle \delta W_{\mathrm{mem}}^\ast \rangle$ are also investigated.
Two different stable modes are controlled by the initial orientation angle $\Psi_0$,
as seen in figure~\ref{fig:map_ori}.

\begin{figure}
  \centering
  \includegraphics[height=5.5cm]{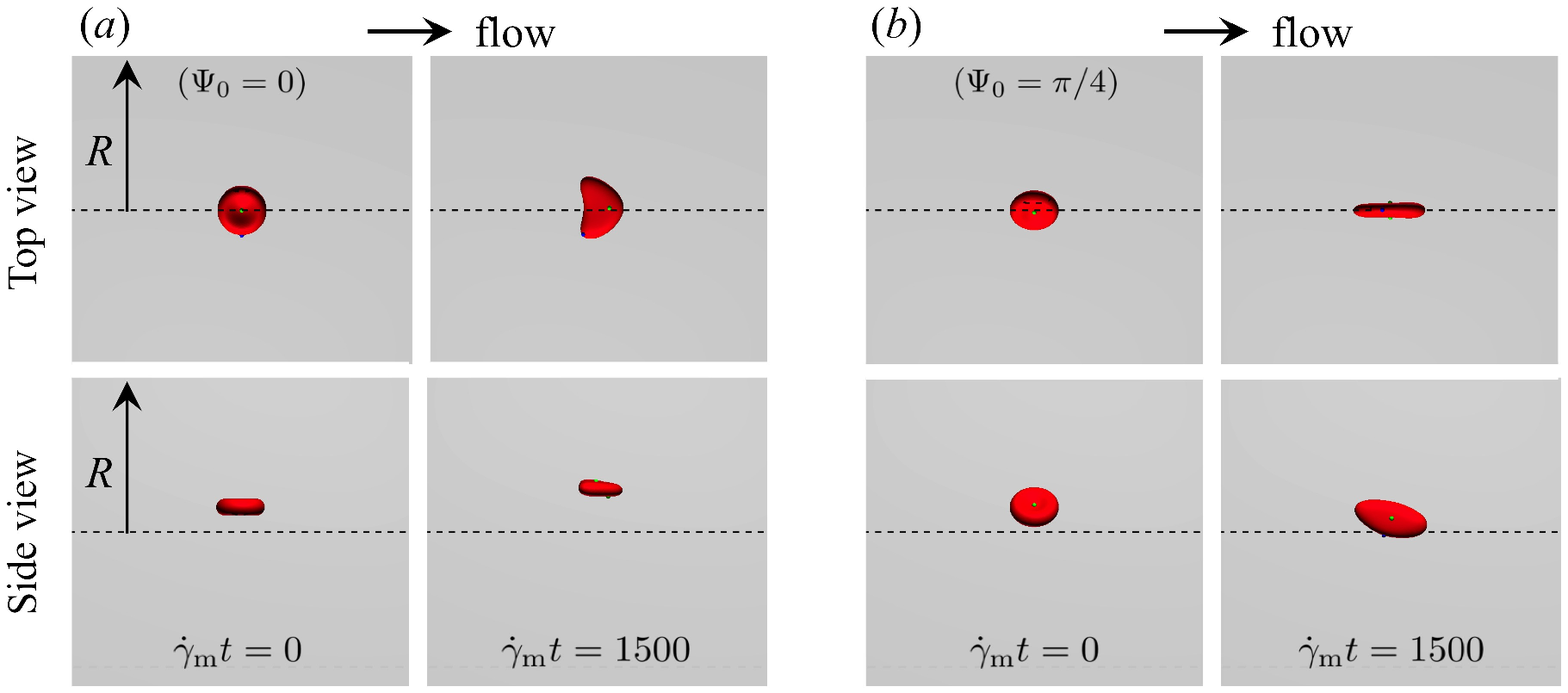}
  \includegraphics[height=5.5cm]{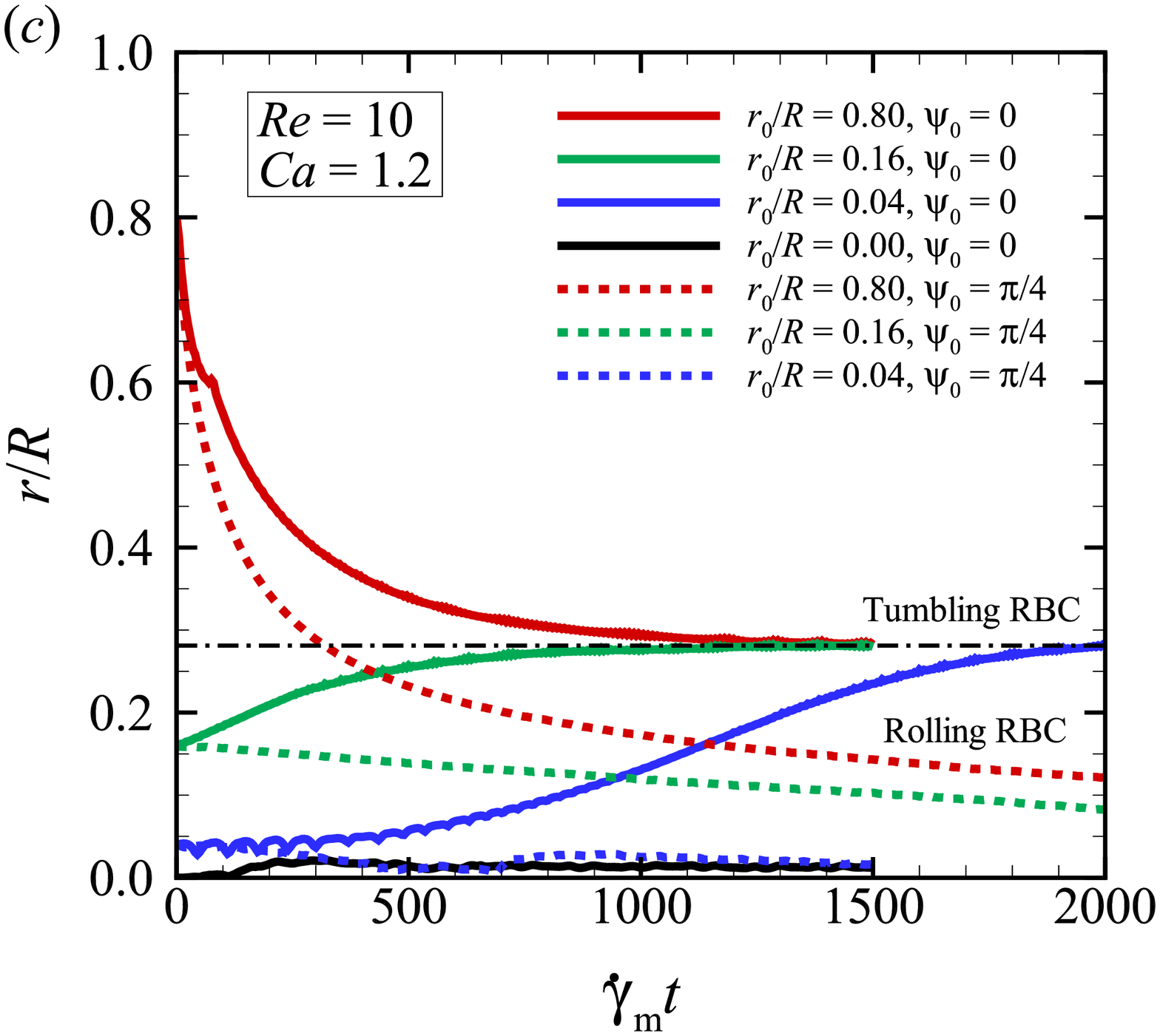}
  \caption{
	  Snapshots of RBCs at initial state ($\dot{\gamma}_\mathrm{m} t$ = 0) and during fully developed tumbling and rolling motions ($\dot{\gamma}_\mathrm{m} t$ = 1500),
	  where the initial orientation angles are set to ($a$) $\Psi_0$ = 0 and ($b$) $\Psi_0$ = $\pi/4$, respectively.
	  The upper and lower snapshots show the top view and side views, respectively.
	  ($c$) Time history of the radial position of RBC centroids for different $r_0/R$ and $\Psi_0$.
	  Solid lines indicate $\Psi_0$ = 0 and dashed lines indicate $\Psi_0$ = $\pi/4$.
	  The dash-dot line represents the equilibrium radial position of tumbling RBCs with $r/R$ = 0.2812.
      	  The results are obtained with $Re$ = 10 and $Ca$ = 1.2.
  }
  \label{fig:re10ca1_2}
\end{figure}

Figures~\ref{fig:re10ca1_2}($a$) and \ref{fig:re10ca1_2}($b$) show snapshots of flowing RBCs subject to the highest $Ca$ (= 1.2) for different initial orientation angles $\Psi_0$ (= 0 and $\pi/4$) at the initial state ($\dot{\gamma}_\mathrm{m} t$ = 0) and the final time point ($\dot{\gamma}_\mathrm{m} t$ = 1500).
The RBC starting from $\Psi_0$ = 0 exhibits a flipping or tumbling motion,
and assumes a flattened croissant-like shape that is convex at the front and concave at the rear (figure~\ref{fig:re10ca1_2}$a$).
This tumbling motion allows the RBC to migrate towards the wall from the initial near-center position $r_0/R$ = 0.16 to $r/R$ = 0.2812 (figure~\ref{fig:re10ca1_2}$c$, green solid line).
The RBCs initially placed near a center positions ($r_0/R$ = 0.04) with $\Psi_0$ = 0 require a long period of time to reach this $r/R$ threshold (= 0.2812) (figure~\ref{fig:re10ca1_2}$c$, solid blue line).
The RBC starting from $\Psi_0 = \pi/4$ exhibits an elongated rolling motion,
as shown in figure~\ref{fig:re10ca1_2}($b$),
and tends to migrate towards the channel center (figure~\ref{fig:re10ca1_2}$c$, green dashed line).
Such equilibrium radial positions are basically independent of the initial radial position $r_0/R$,
except for the case of $r_0/R$ = 0,
where the RBC starting from $\Psi_0$ = 0 remains almost on the channel axis, $O(r/R)$ = 10$^{-3}$,
without an obvious tumbling motion (figure~\ref{fig:re10ca1_2}$c$, black solid line).
Travel on the channel axis is also observed in the case with $r_0/R$ = 0.04 and $\Psi_0$ = $\pi$/4 (figure~\ref{fig:re10ca1_2}$c$, blue dashed line).
Considering a linear estimation for the speed of axial migration,
corresponding to the gradient of the time history of radial positions of RBC centroids,
using data between $\dot{\gamma}_\mathrm{m} t$ = 1500 and 2000,
rolling RBCs starting from $r_0/R$ = 0.8 and 0.16 will reach the near-center position $O(r/R) \leq$ 10$^{-2}$ at $\dot{\gamma}_\mathrm{m}t$ = 4500 and 3800, respectively.

\begin{figure}
  \centering
  \includegraphics[height=5.5cm]{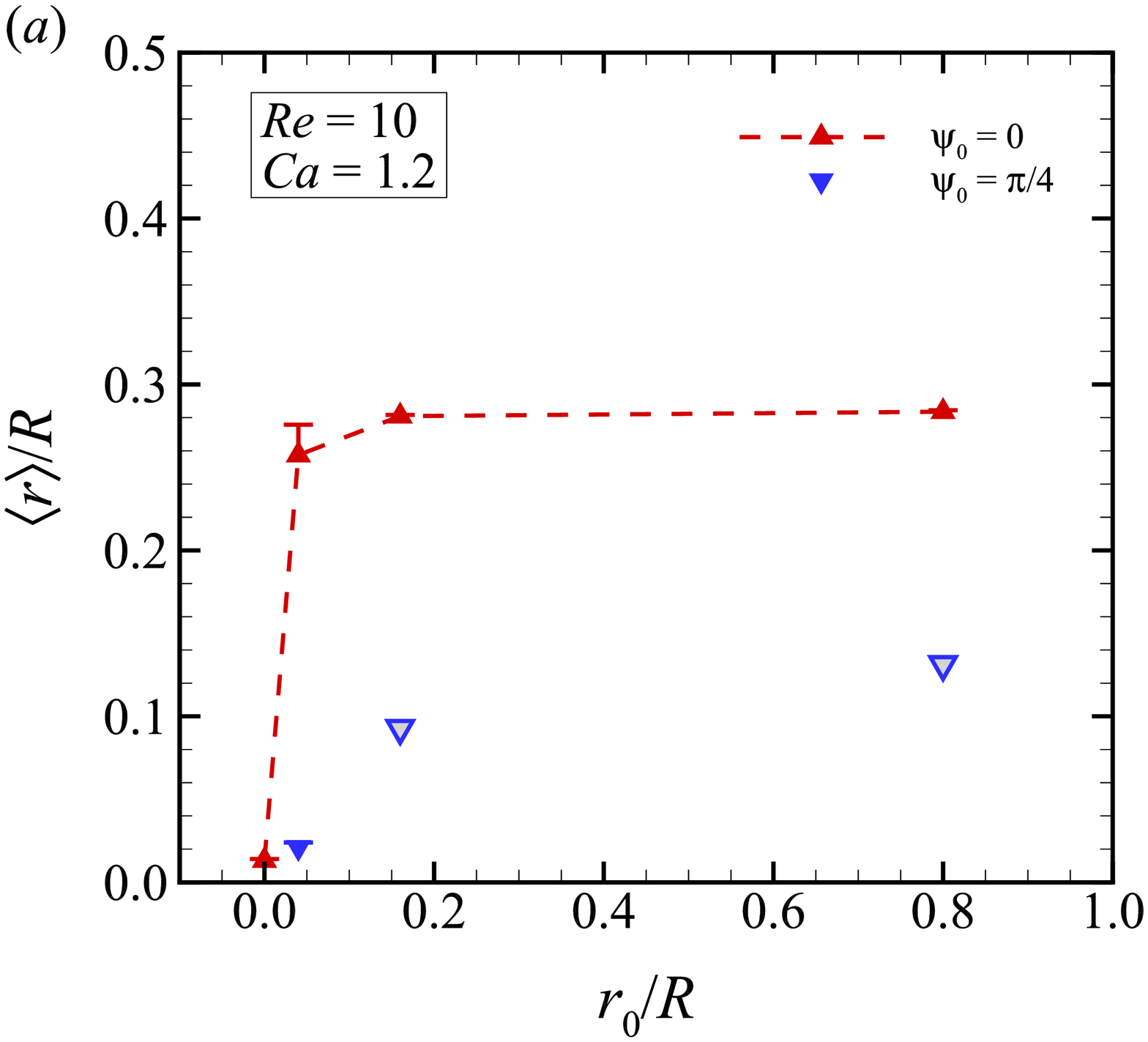}
  \includegraphics[height=5.5cm]{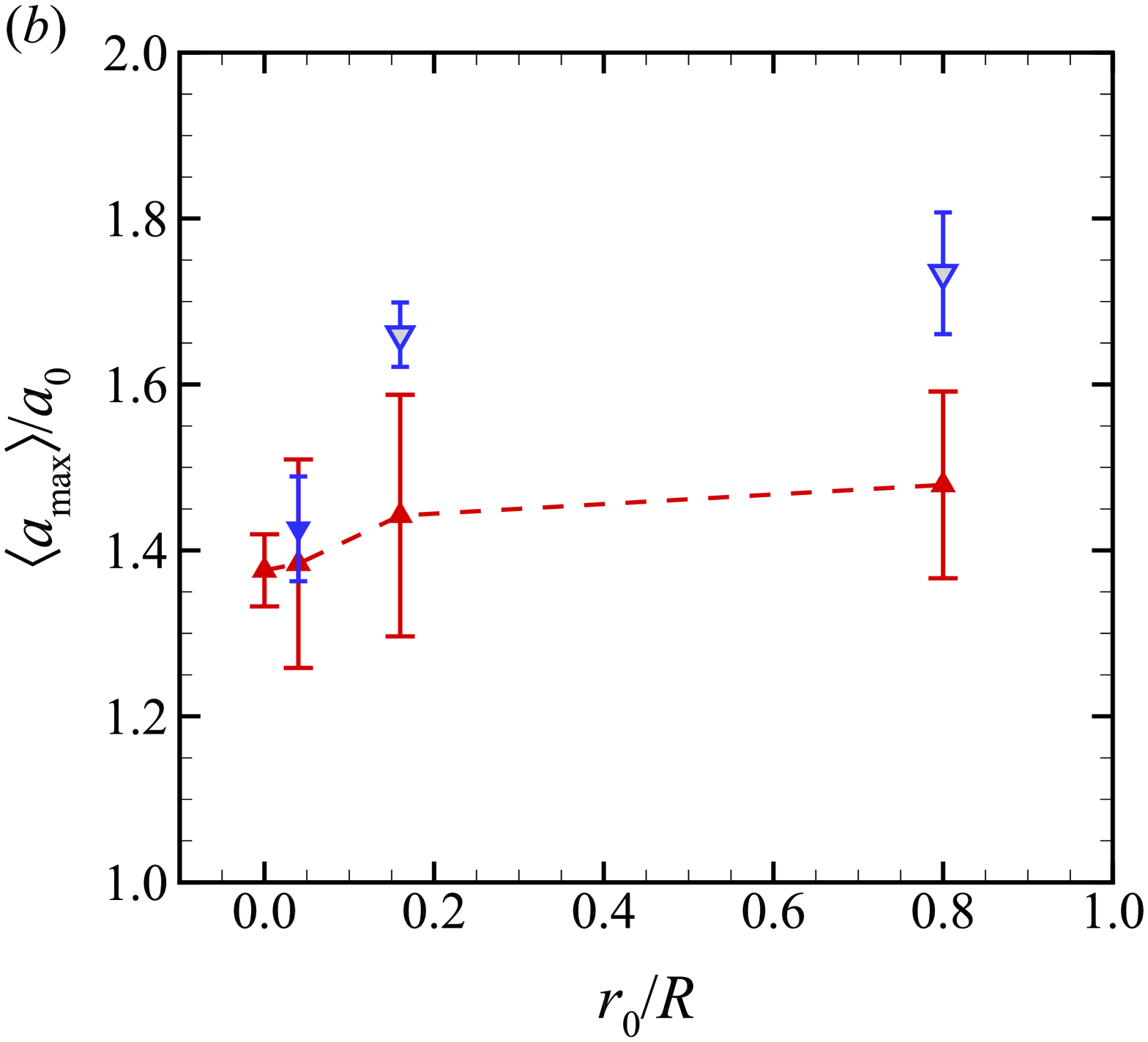}
  \includegraphics[height=5.5cm]{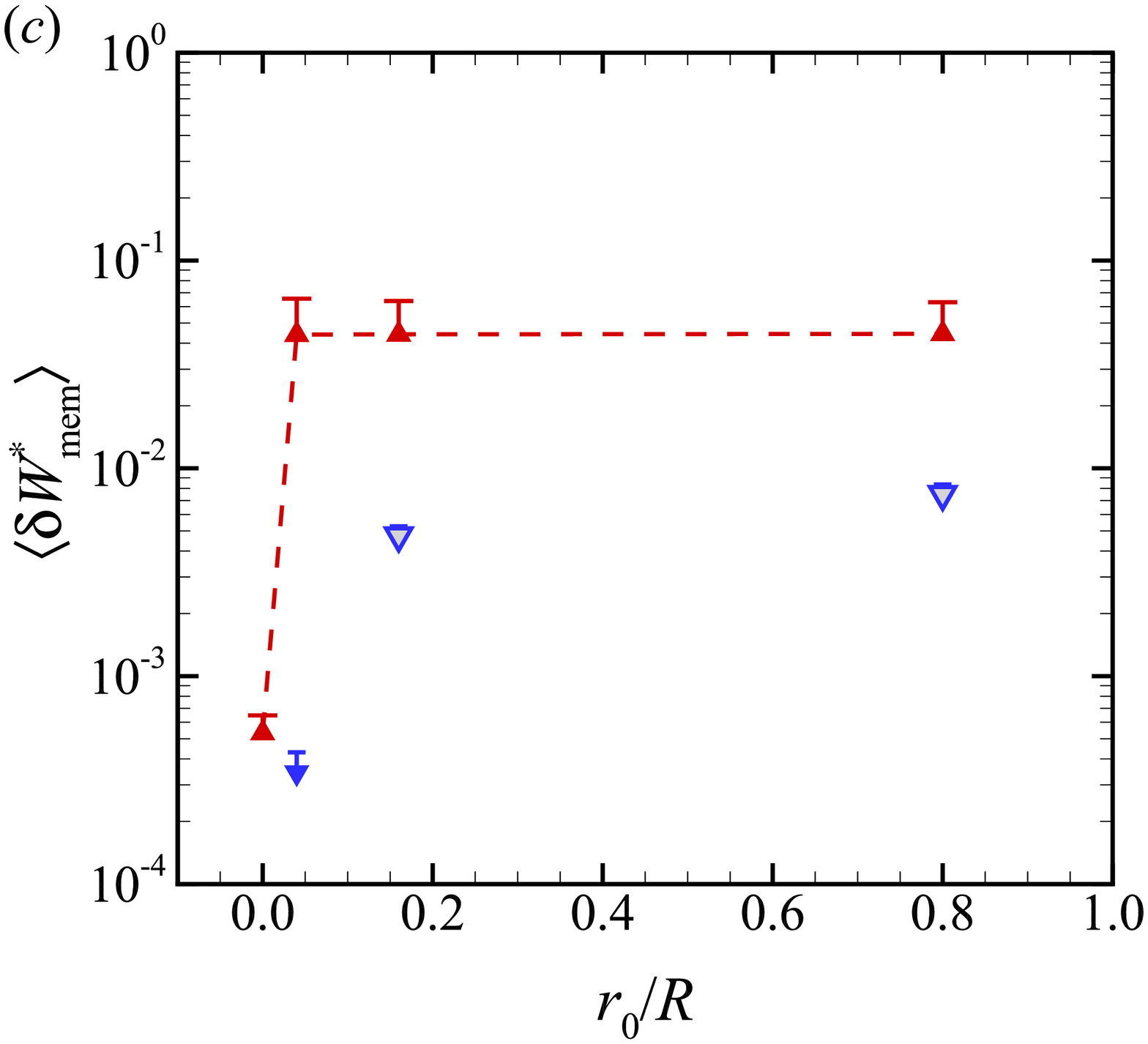}
  \includegraphics[height=5.5cm]{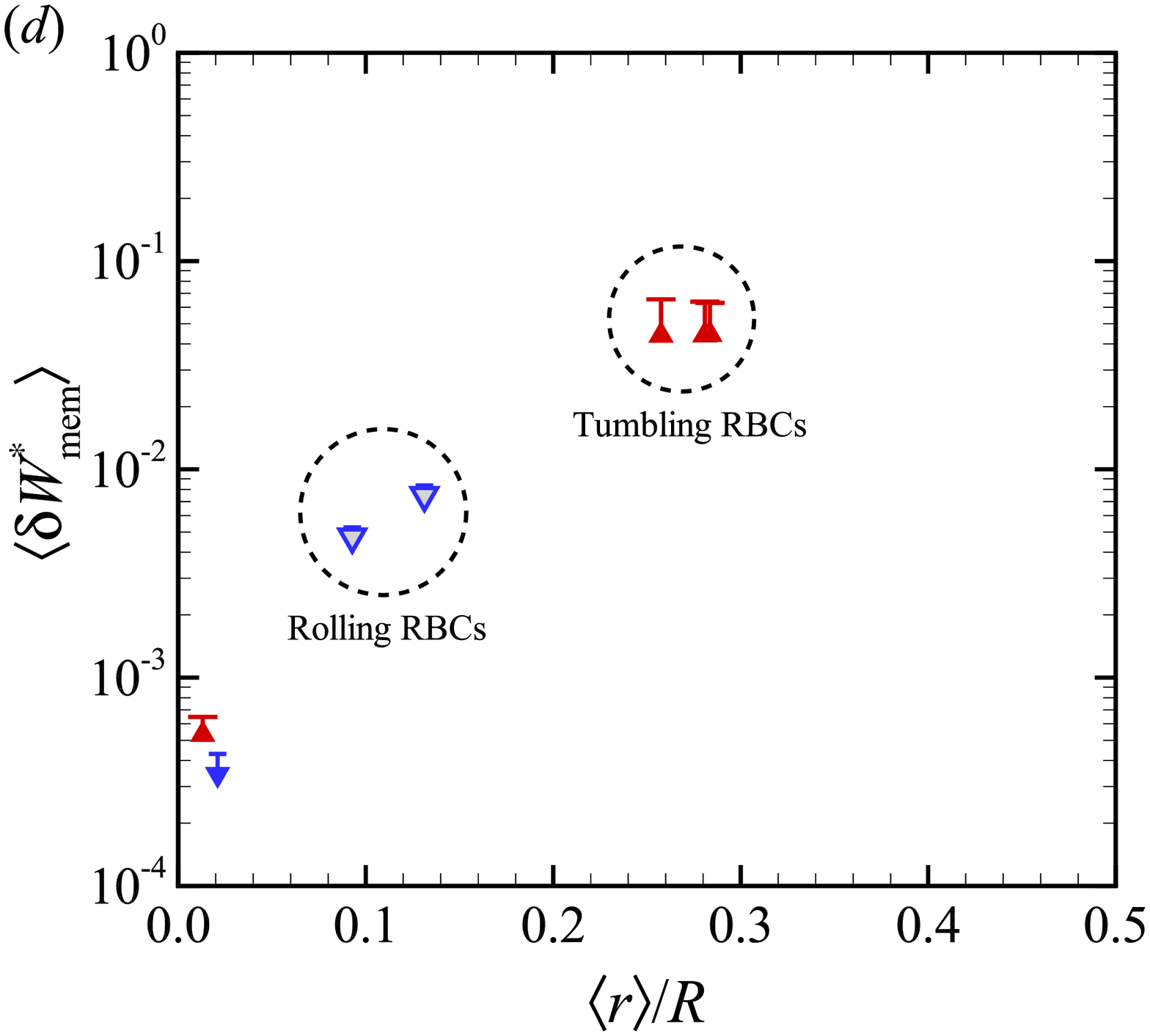}
  \caption{
	  ($a$) Time average of the radial position $\langle r \rangle/R$ of a RBC centroid,
	  ($b$) the maximum length $\langle a_{\mathrm{max}} \rangle/a_0$ of a deformed RBC, and
	  ($c$) powers $\langle \delta W_{\mathrm{mem}}^\ast \rangle$ associated with membrane deformations.
	  All are shown as a function of the initial radial position $r_0/R$ and initial orientation angles $\Psi_0$ (= 0 and $\pi/4$).
	  ($d$) Replotted data of $\langle \delta W_{\mathrm{mem}}^\ast \rangle$ as a function of equilibrium radial position $\langle r \rangle/R$.
  	  Rolling RBCs starting from $\Psi_0$ = $\pi$/4 and $r_0/R \geq$ 0.16 are in transit (still migrating toward the channel center),
	  and are represented by light gray dots.
	  All results are obtained with $Re$ = 10 and $Ca$ = 1.2.
  }
  \label{fig:effect_r0}
\end{figure}

Figure~\ref{fig:effect_r0}($a$) shows the time average of the radial position $\langle r \rangle/R$ as a function of the initial radial position $r_0/R$ for different initial orientation angles $\Psi_0$ (= 0, and $\pi$/4).
As described above,
tumbling RBCs reach the threshold value, $r/R$ = 0.2812,
while rolling RBCs exhibit axial migration (figure~\ref{fig:effect_r0}$a$).
Since rolling RBCs are elongated in the flow direction,
their maximal radius $a_\mathrm{max}$ in the deformed shape tends to be greater than that of tumbling RBCs (figure~\ref{fig:effect_r0}$b$).
Due to the large cyclic extension of tumbling RBCs,
as shown in figure~\ref{fig:map_ori}($b$),
the time-dependent fluctuation in $a_\mathrm{max}/a_0$ is greater in tumbling RBCs than in rolling ones.

Figure~\ref{fig:effect_r0}($c$) shows the powers $\langle \delta W_\mathrm{mem}^\ast \rangle$ as a function of the initial radial position $r_0/R$ for different initial orientation angles $\Psi_0$.
Since RBCs flowing near the channel axis are associated with low energy expenditure independently of $\Psi_0$,
it is expected that rolling RBCs with 0.16 $\leq r_0/R \leq$ 0.8 (light gray dots in figure~\ref{fig:effect_r0}$c$) will also reach a small order of magnitude of the powers $O(\langle \delta W_\mathrm{mem}^\ast \rangle) \leq$ 10$^{-3}$.
The results of $\langle \delta W_\mathrm{mem}^\ast \rangle$ are replotted as a function of $\langle r \rangle/R$ (figure~\ref{fig:effect_r0}$d$).
The result suggests that the orders of magnitude of the powers $\langle \delta W_\mathrm{mem}^\ast \rangle$ correlates with the (equilibrium) radial position,
which in turn is associated with stable flow mode.

\subsection{Effect of $Re$ on equilibrium radial position}
\begin{figure}
  \centering
  \includegraphics[height=4cm]{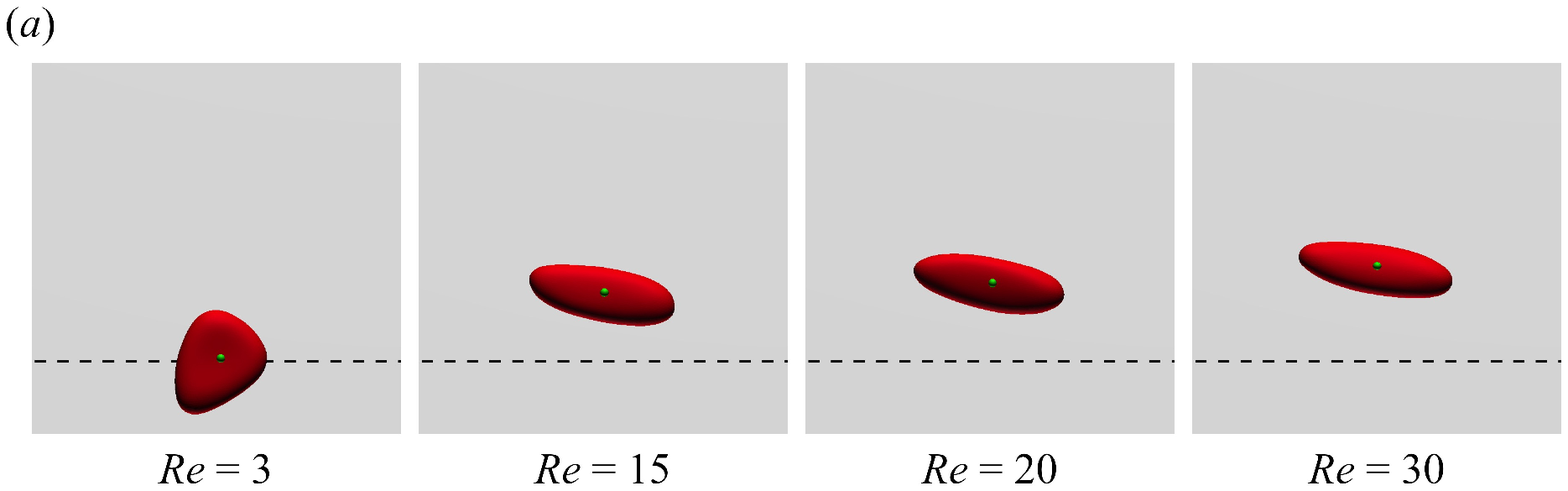}  
  \includegraphics[height=5.5cm]{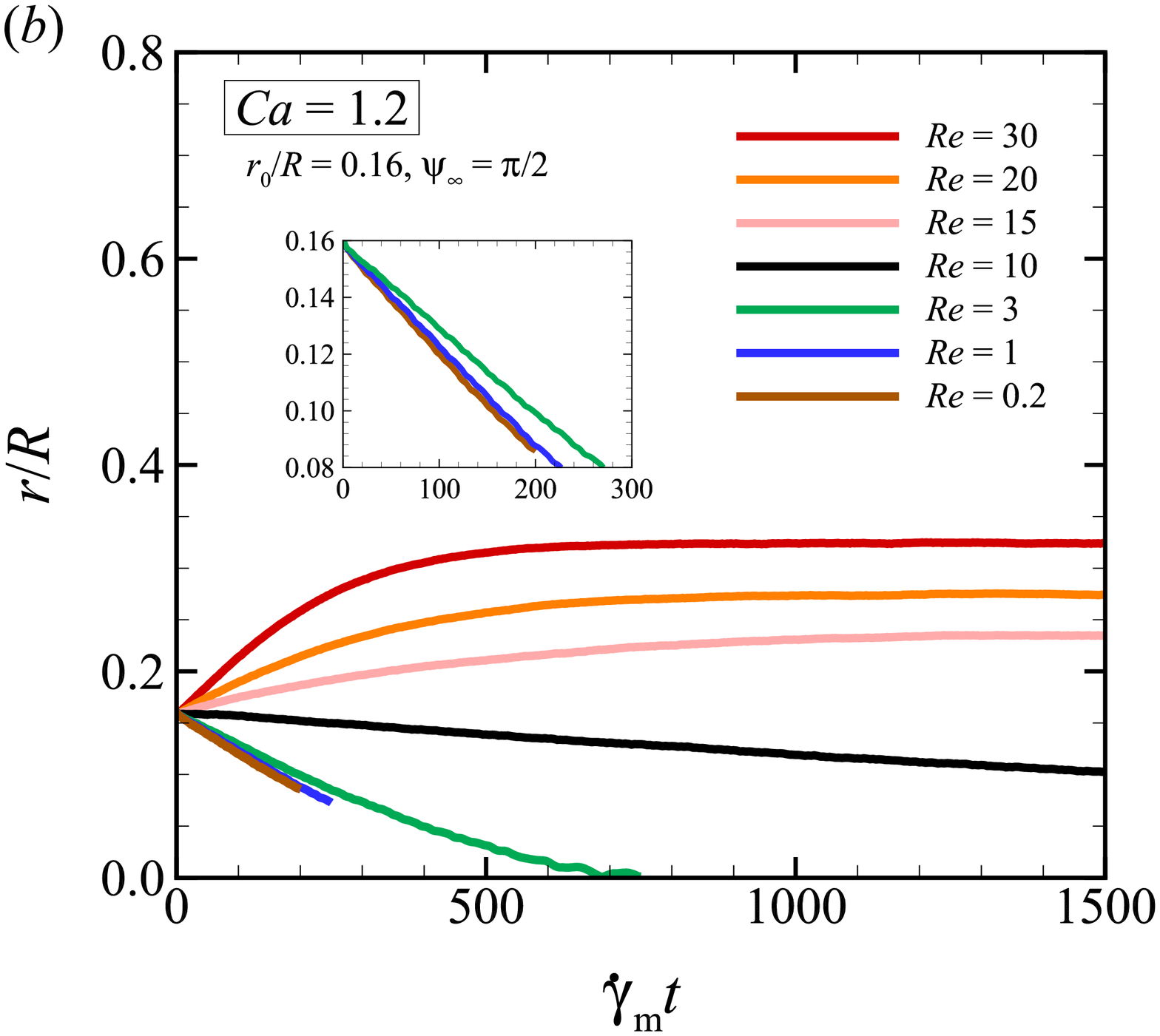}
  \caption{
	  ($a$) Snapshots of RBCs subject to $Ca$ = 1.2 for different $Re$ at the equilibrium radial position: $\dot{\gamma}_\mathrm{m} t$ = 750 for $Re$ = 3,
	  and $\dot{\gamma}_\mathrm{m} t$ = 1500 for $Re$ = 15, 20, and 30.
	  ($b$) Time history of radial positions of RBC centroids for different $Re$.
  	  All RBCs exhibit rolling motion ($|\Psi_\infty| \sim \pi$/2).
	  The results are obtained with $Ca$ = 1.2 and $r_0/R$ = 0.16.
  }
  \label{fig:timehist_ca1_2_re}
\end{figure}

The effect of $Re$ on the equilibrium radial positions of RBC centroids is also investigated for specific $Ca$ = 1.2 and initial radial position $r_0/R$ = 0.16.
Representative snapshots of RBCs at each equilibrium radial position are shown in figure~\ref{fig:timehist_ca1_2_re}($a$).
Off-centered rolling of RBCs is clearly seen for $Re \geq$ 15,
while the axially migrated RBC under $Re$ = 3 does not exhibit a characteristic flow mode because of the low shear rate region (figure~\ref{fig:timehist_ca1_2_re}$a$).
Figure~\ref{fig:timehist_ca1_2_re}($b$) shows the time history of RBC centroids $r/R$ for different $Re$.
The inertial migration of rolling RBCs subject to $Ca$ = 1.2 requires at least $Re \geq$ 15 (figure~\ref{fig:timehist_ca1_2_re}$b$).
The speed of axial migration increases as $Re$ ($\leq$ 10) decreases (figure~\ref{fig:timehist_ca1_2_re}$b$).
For RBCs at lower $Re \leq$ 1,
axial speeds linearly estimated using data between $\dot{\gamma}_\mathrm{m} t$ = 0 and 200 predict axial migration ($O(\langle r \rangle/R) \leq$ 10$^{-2}$) within $\dot{\gamma}_\mathrm{m}t$ = 500.
Complete axial migration at such low $Re$ cannot be practically confirmed due to heavy computational load.
Instead, we perform a simulation at $Re$ = 0.2 in a smaller channel with $D$ = 20 $\mu$m,
and find that RBCs exhibit axial migration regardless of $Ca$ (see figure~\ref{fig:d20re02}, in Appendix~\ref{appA3}).
In the time-averaged results,
data are shown only for cases in which RBCs have reached each equilibrium radial position (i.e., $Re$ = 3, 15, 20, and 30).

\begin{figure}
  \centering
  \includegraphics[height=5.5cm]{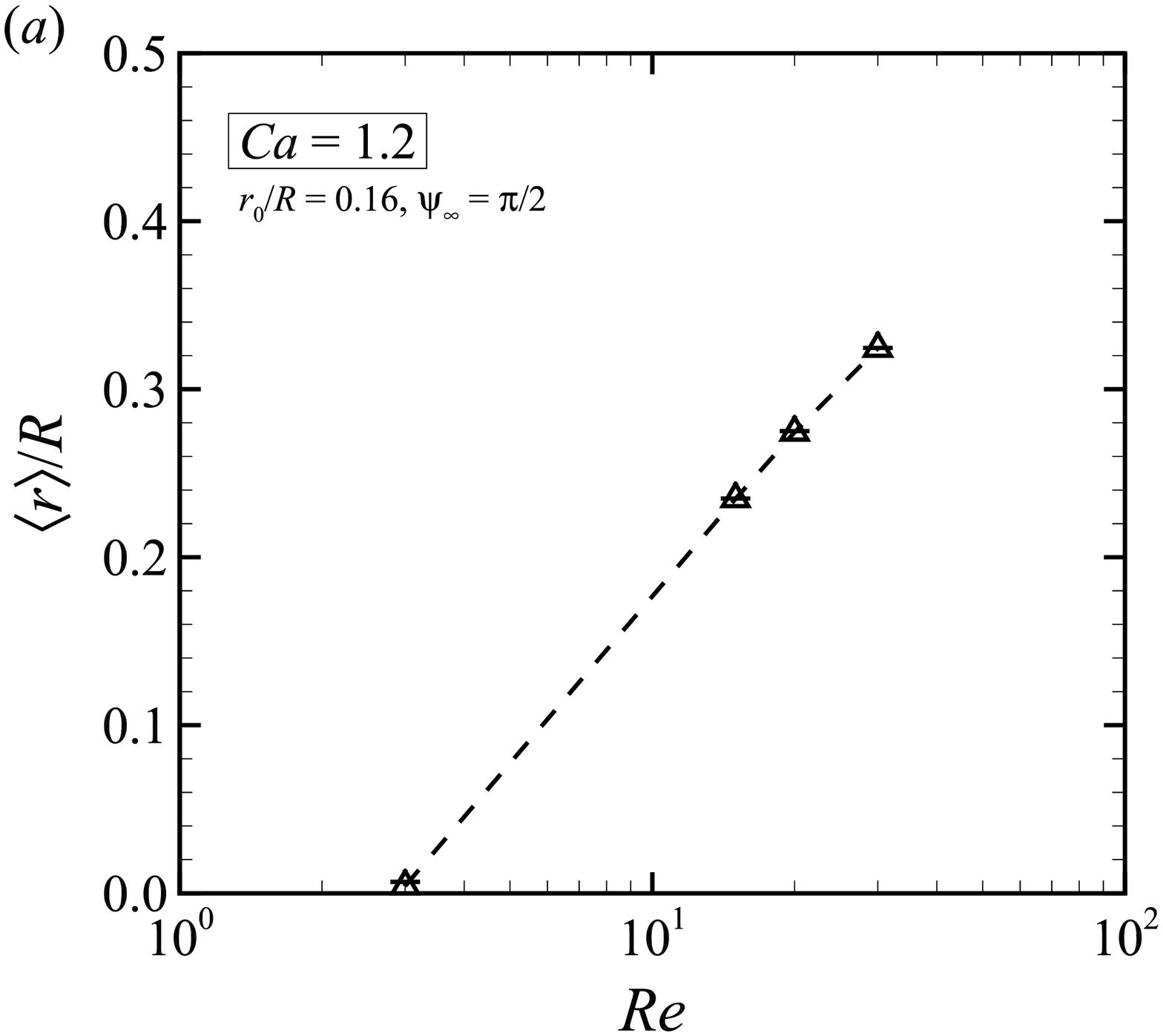}
  \includegraphics[height=5.5cm]{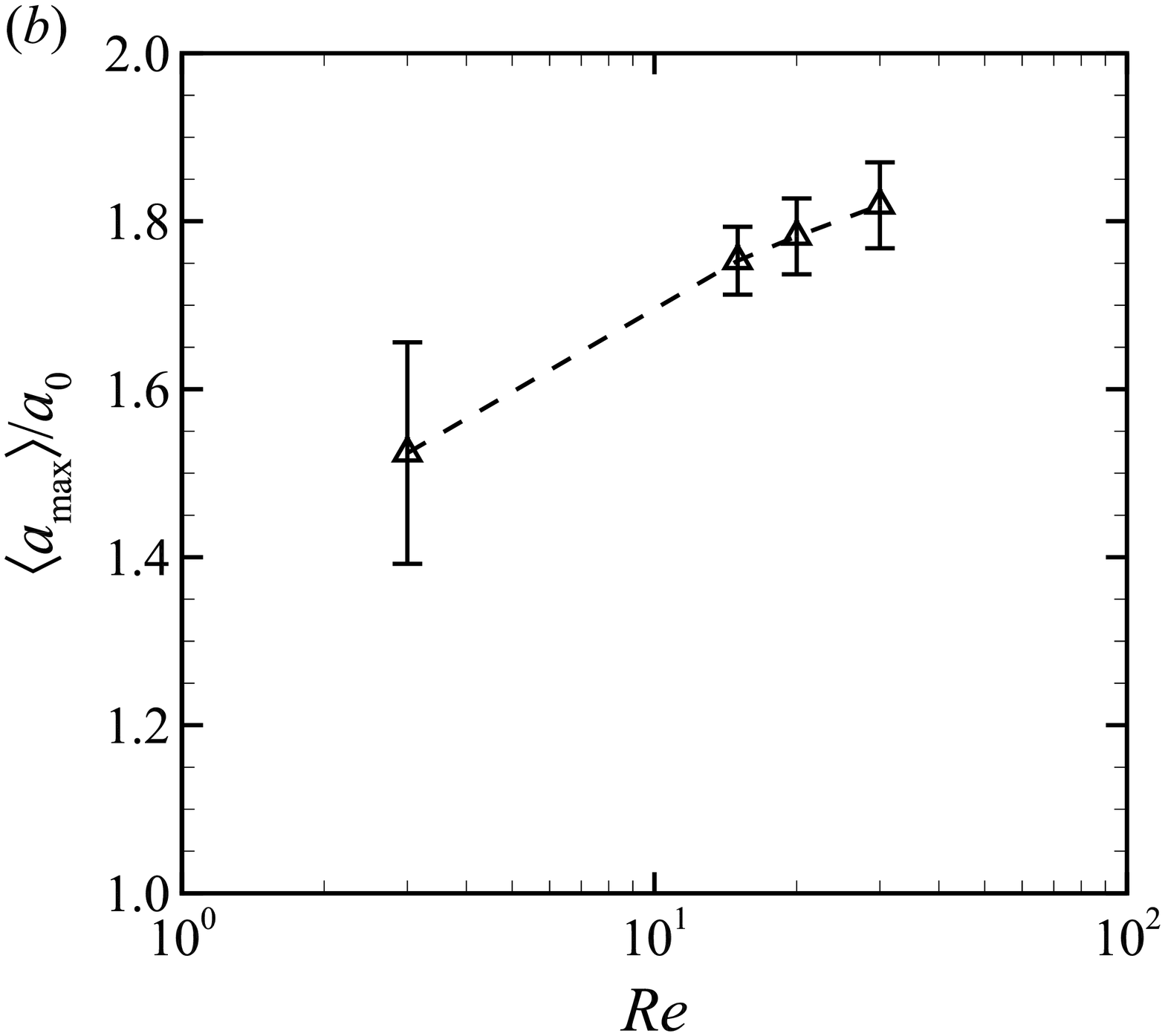}
  \includegraphics[height=5.5cm]{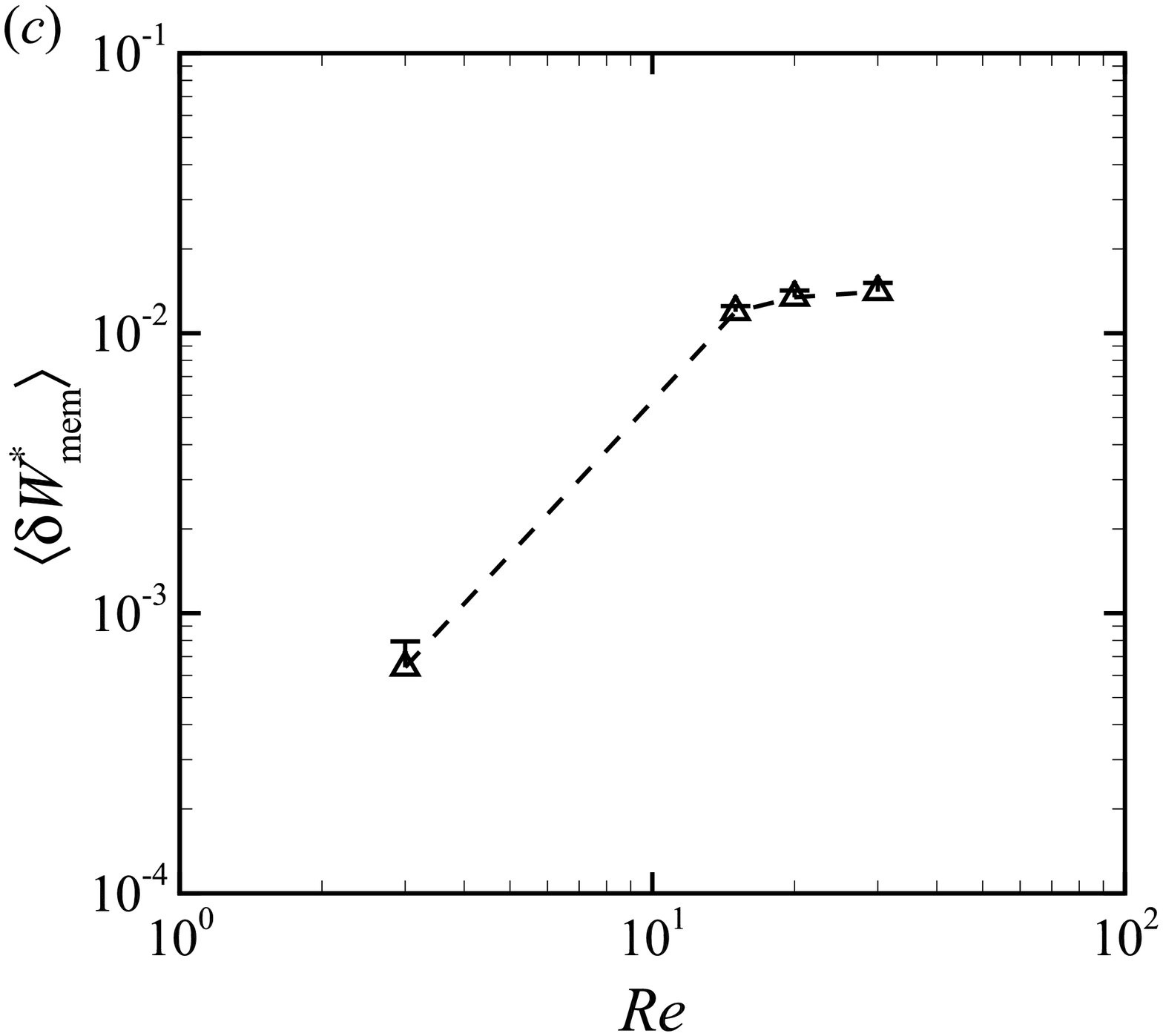}
  \includegraphics[height=5.5cm]{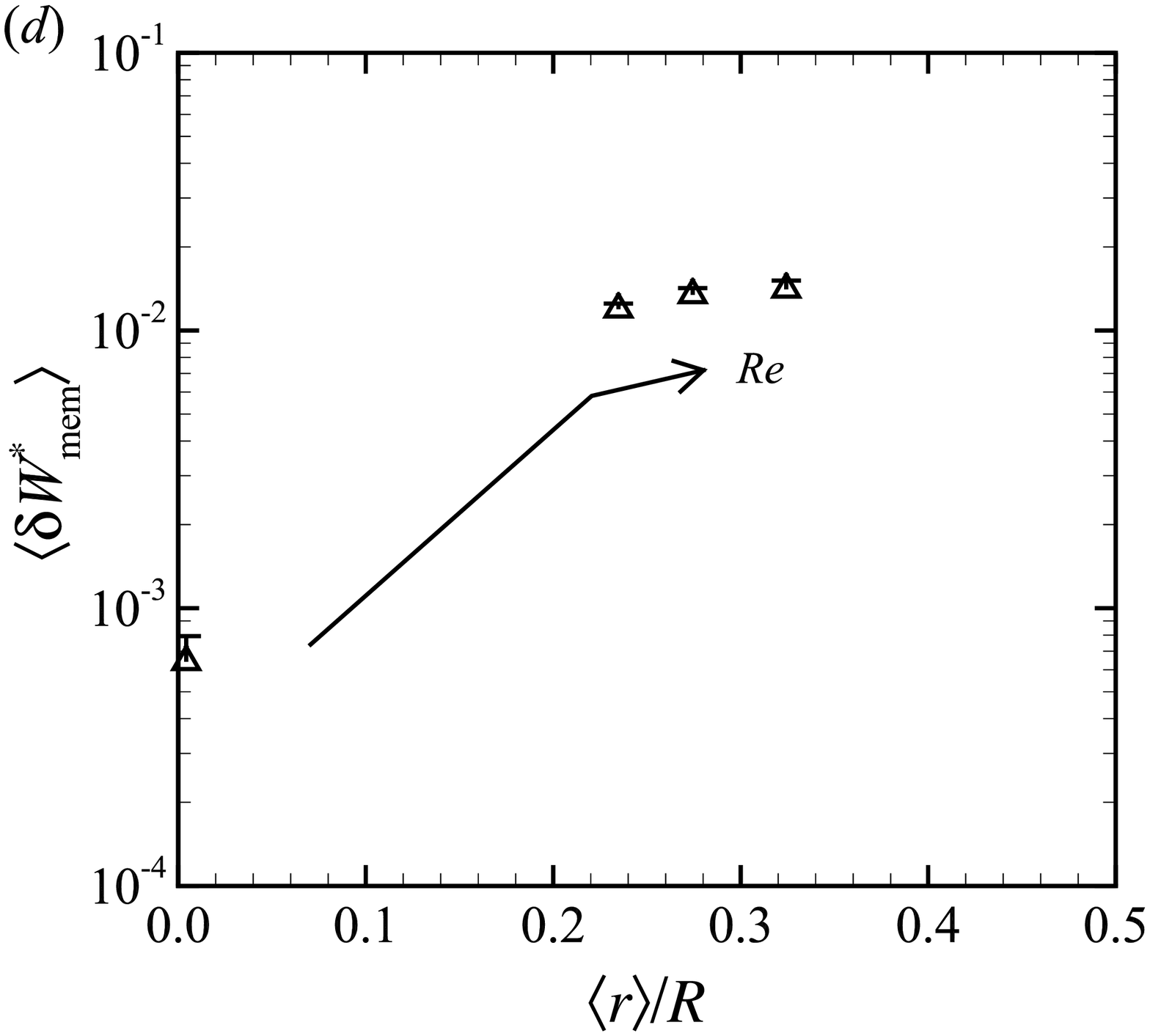}
  \caption{
	  ($a$) Time average of the radial position $\langle r \rangle/R$ of an RBC centroid,
	  ($b$) the maximum length $\langle a_{\mathrm{max}} \rangle/a_0$ of a deformed RBC, and
	  ($c$) powers $\langle \delta W_{\mathrm{mem}}^\ast \rangle$ associated with membrane deformations.
	   All are shown as a function of $Re$.
	  ($d$) Replotted data of $\langle \delta W_{\mathrm{mem}}^\ast \rangle$ as a function of equilibrium radial position $\langle r \rangle/R$ for different $Re$.
	  Results are obtained with $Ca$ = 1.2 and $r_0/R$ = 0.16.
  }
  \label{fig:effect_re}
\end{figure}

Figure~\ref{fig:effect_re}($a$) shows the time average of the radial position $\langle r \rangle/R$ as a function of $Re$.
RBCs flowing at large $\langle r \rangle/R$ experience high shear stress,
resulting in large deformation $a_{\mathrm{max}}/a_0$ as shown in figure~\ref{fig:effect_re}($b$).

Figure~\ref{fig:effect_re}($c$) shows the powers $\langle \delta W_{\mathrm{mem}}^\ast \rangle$ as a function of $Re$.
Although axially migrated RBCs are associated with low energy expenditure $O(\langle \delta W_{\mathrm{mem}}^\ast \rangle)$ = 10$^{-3}$,
inertially migrated RBCs are associated with high energy expenditure $O(\langle \delta W_{\mathrm{mem}}^\ast \rangle)$ = 10$^{-2}$ (figure~\ref{fig:effect_re}$c$),
which is consistent with the results in figure~\ref{fig:effect_r0}($c$).
The relationship between $\langle \delta W_\mathrm{mem}^\ast \rangle$ and $Re$ remains the same even when the data are replotted as a function of $\langle r \rangle/R$,
where $\langle \delta W_\mathrm{mem}^\ast \rangle$ increases with increased $Re$ or $\langle r \rangle/R$ increases (figure~\ref{fig:effect_re}$d$).

\subsection{Effect of viscosity ratio $\lambda$ on equilibrium radial position}
\begin{figure}
  \centering
  \includegraphics[height=5.5cm]{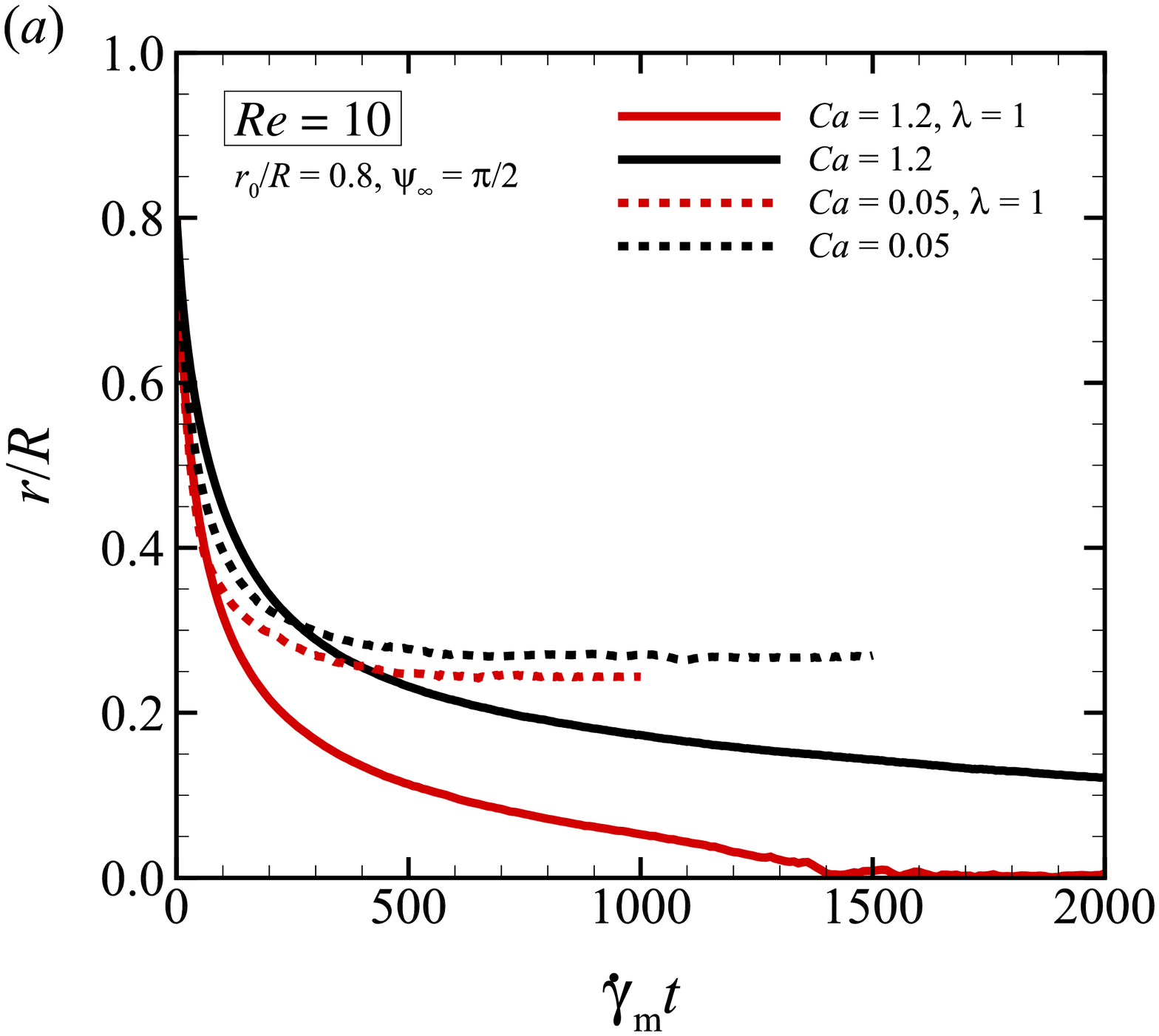}
  \includegraphics[height=5.5cm]{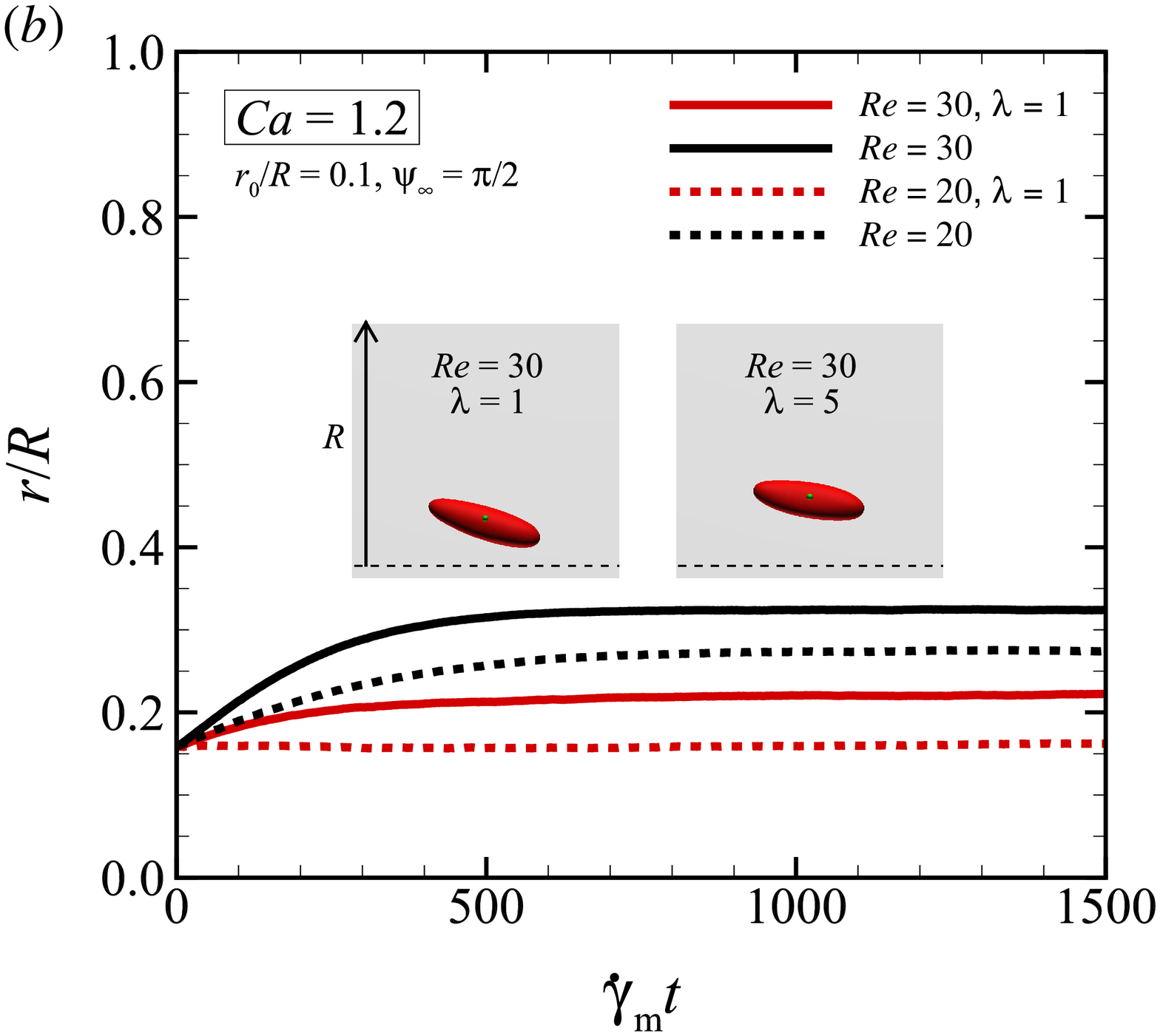}
  \includegraphics[height=5.5cm]{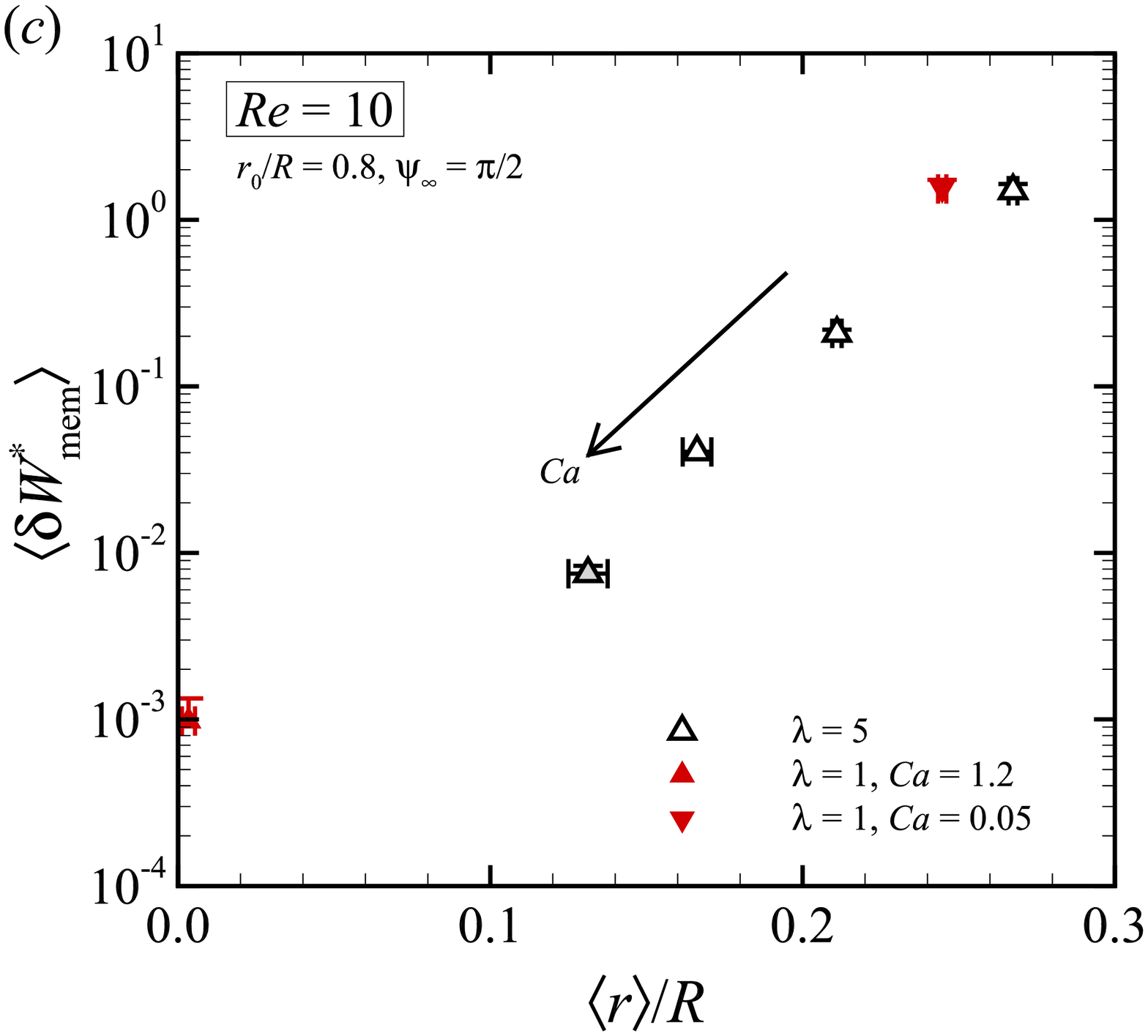}
  \includegraphics[height=5.5cm]{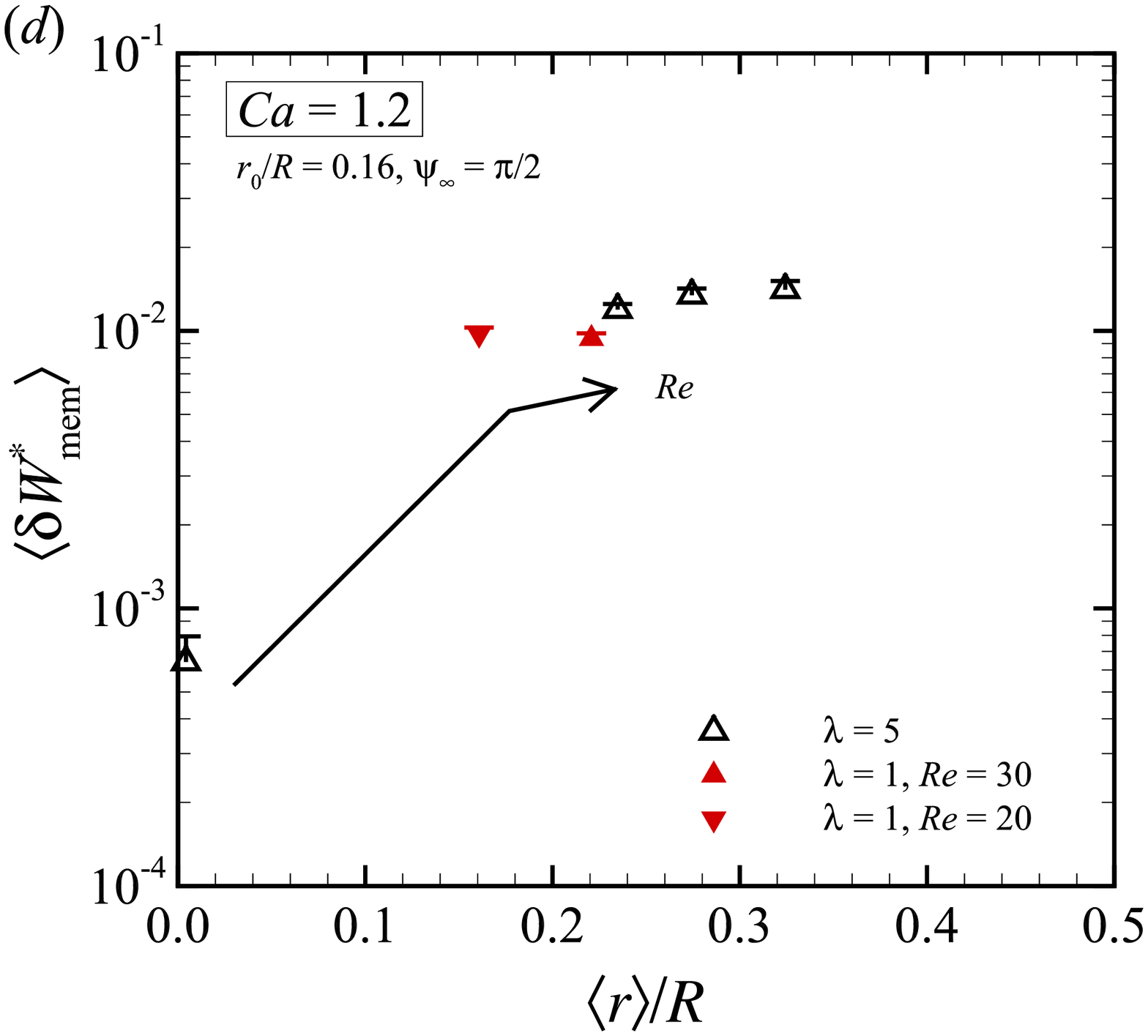}
  \caption{
	  Time history of the radial position of an RBC centroid for different ($a$) $Ca$ and ($b$) $Re$,
	  where the results obtained with $\lambda$ = 1 are added in figure~\ref{fig:timehist_re10_ca}($b$) and figure~\ref{fig:timehist_ca1_2_re}($b$).
	  Insets show snapshots of a flowing RBC at each equilibrium position for $Re$ = 30 and $\lambda$ = 1 and 5.
  	  Replotted data of $\langle \delta W_{\mathrm{mem}}^\ast \rangle$ as a function of equilibrium radial position $\langle r \rangle/R$ for ($c$) different $Ca$ and ($d$) $Re$.
	  All RBCs exhibit stable rolling motion ($|\Psi_\infty| \sim \pi$/2).
  }
  \label{fig:effect_lam}
\end{figure}

To clarify the impact of the physiologically relevant viscosity ratio $\lambda$ (= 5, in this study) on the equilibrium radial positions of RBC centroids,
we simulate the behaviour of RBCs with $\lambda$ being unity.
The time histories of radial position $r/R$ obtained with $\lambda$ = 1 are added in figure~\ref{fig:timehist_re10_ca}($b$) and figure~\ref{fig:timehist_ca1_2_re}($b$),
and are shown in figures~\ref{fig:effect_lam}($a$) and \ref{fig:effect_lam}($b$), respectively.
When $\lambda$ decreases from 5 to 1,
the RBCs under $Re$ = 10 immediately migrate towards the channel center,
and the equilibrium radial position is decreased at each $Ca$ (= 0.05 and 1.2)(figure~\ref{fig:effect_lam}$a$).
In particular, at the highest $Ca$ = 1.2, the RBC exhibits complete axial migration with $\dot{\gamma}_\mathrm{m} t$ = 1500.
The decrease in the equilibrium radial positions of RBC centroids is consistent even for large $Re$,
as shown in figure~\ref{fig:effect_lam}($b$).
However the RBC subject to $Ca$ = 1.2 still exhibits inertial migration,
at least for $Re \geq$ 20,
even for $\lambda$ = 1 (figure~\ref{fig:effect_lam}$b$).
For the additional runs with $\lambda$ = 1,
the stable flow mode remains the rolling motion.
Overall, the low $\lambda$ (= 1) condition impedes inertial migration (figures~\ref{fig:effect_lam}$a$ and \ref{fig:effect_lam}$b$).

The relationship between $\langle r \rangle/R$ and $\langle \delta W_\mathrm{mem} \rangle$ obtained with $\lambda$ = 1 is superimposed on the results for $\lambda$ = 5 (figures~\ref{fig:effect_ca}$d$ and figure~\ref{fig:effect_re}$d$),
and the results are plotted on an estimated curve obtained with $\lambda$ = 5,
as shown in figures~\ref{fig:effect_lam}($c$) and \ref{fig:effect_lam}($d$).
The tendency remains the same even at low $\lambda$ (= 1).

\subsection{Equilibrium radial position of a spherical capsule}
\begin{figure}
  \centering
  \includegraphics[height=3.5cm]{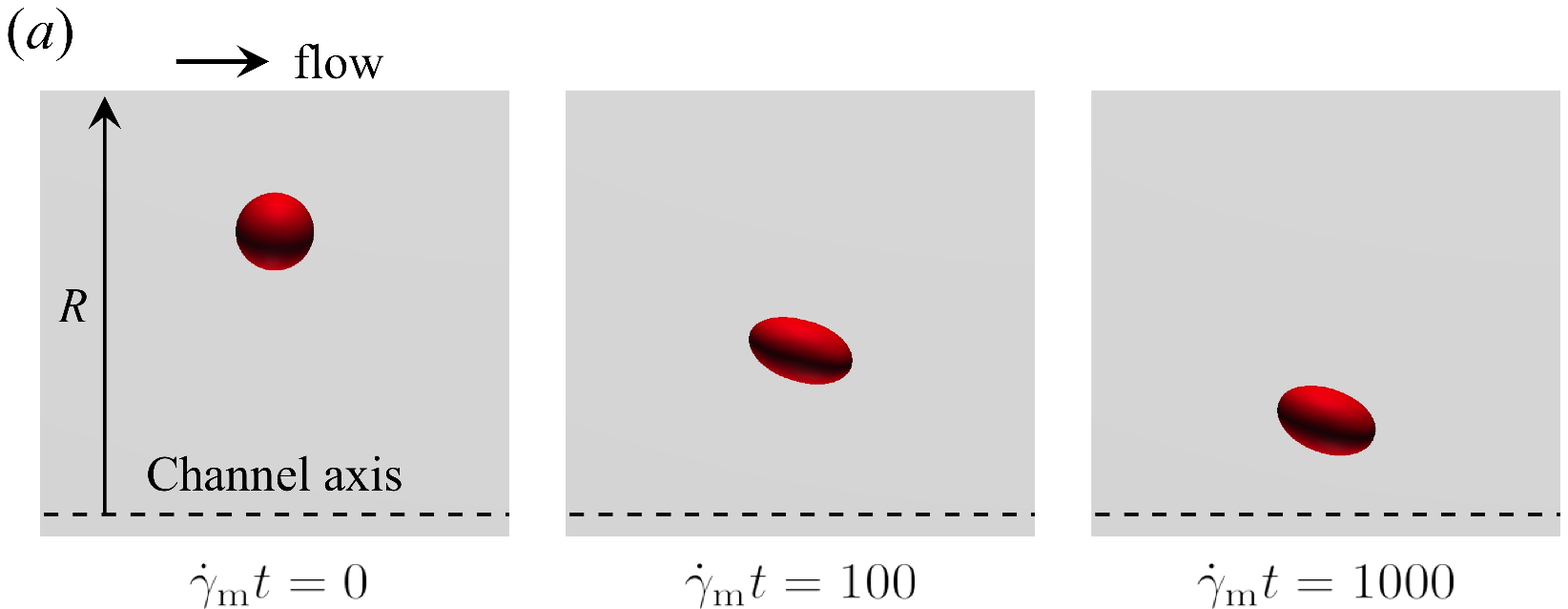}
  \includegraphics[height=5.5cm]{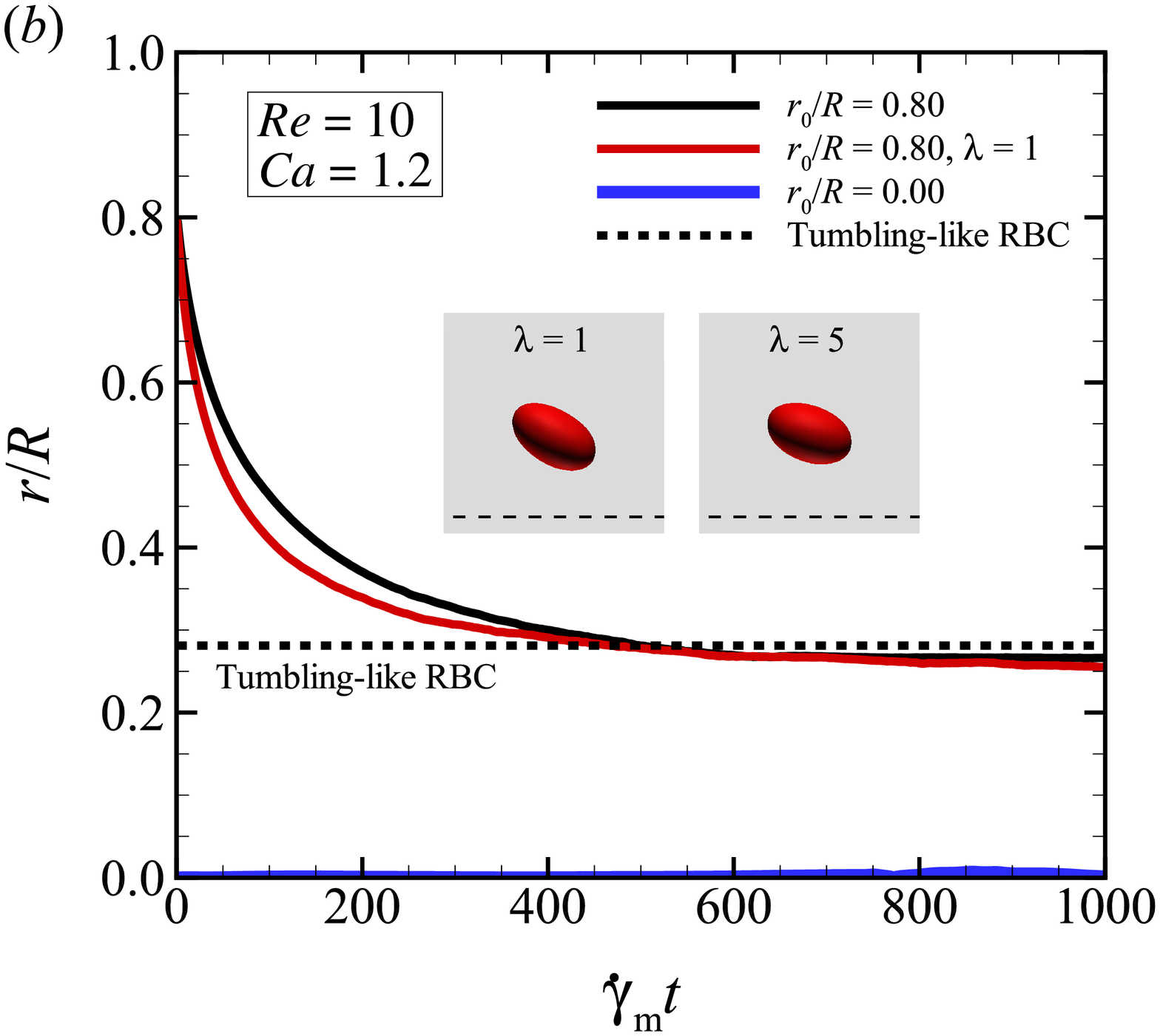}
  \caption{
  	  ($a$) Snapshots of a spherical capsule with $\lambda$ = 5 initially placed at $r_0/R$ = 0.8 for $Ca$ = 1.2 and $Re$ = 10 at specific time points $\dot{\gamma}_\mathrm{m} t$ = 0 ({\it left}),
	  $\dot{\gamma}_\mathrm{m} t$ = 100 ({\it center}),
	  and $\dot{\gamma}_\mathrm{m} t$ = 1000 ({\it right}).
	  ($b$) Time history of the radial positions of spherical capsules for different $r_0/R$ (= 0 and 0.8) and different $\lambda$ = 1 and 5.
	  Insets show snapshots of spherical capsules at each equilibrium position.
	  The equilibrium radial position ($r/R$ = 0.2812) obtained with a tumbling RBC with $\lambda$ = 5 (figure~\ref{fig:re10ca1_2}$b$) is also displayed as black dashed line.
	  }
  \label{fig:re10ca1_2_sphere}
\end{figure}

To clarify whether the equilibrium radial positions of a biconcave capsule with the initial position $r_0$ shown in figure~\ref{fig:re10ca1_2}($c$) can also be predicted using a spherical capsule,
simulations are performed with a spherical capsule whose radius is defined to yield the same volume as the biconcave capsule (i.e., $a_\mathrm{sphere}$ = 2.71 $\mu$m).
Representative snapshots of the spherical capsule starting with $r_0/R$ = 0.8 are shown in figure~\ref{fig:re10ca1_2_sphere}($a$),
where the viscosity ratio $\lambda$ is set to be the same as with the biconcave capsule ($\lambda$ = 5).
The capsule gradually migrates toward the center,
exhibiting a tank-treading motion,
and reaches $r/R$ = 0.2665 at $\dot{\gamma}_\mathrm{m} t$ = 1000,
which is almost the same value as that obtained with the tumbling motion of the biconcave capsule ($r/R$ = 0.2812, figure~\ref{fig:re10ca1_2}$c$).
A spherical capsule initially placed on the channel axis $r_0/R$ = 0 remains in the initial radial position,
which is consistent with the result obtained with a biconcave capsule (figure~\ref{fig:re10ca1_2}$c$).
The spherical capsule with $\lambda$ = 1 reaches $r/R$ = 0.2552 at $\dot{\gamma}_\mathrm{m} t$ = 1000,
which is still close to the threshold obtained with the RBC (figure~\ref{fig:re10ca1_2_sphere}$b$).

\section{Discussion}
In contrast to a large number of previous studies of RBC flow mode especially under shear flow,
little is known about inertial migration of non-spherical (biconcave) capsule in a channel flow.
Furthermore, it is still uncertain about the relationship between those stable flow modes and equilibrium radial position of an RBC centroid on cross-sectional area of the channel.
We address these issues by numerical simulations in present study.

The dynamics of single RBCs has been well investigated experimentally in particular under simple shear flow fields.
For instance, RBCs subjected to a low shear rate exhibit rigid-body-like flipping, the so-called tumbling motion~\citep{Schmid-Schonbein1969, Fischer2004, Dupire2010} and wheel-like rotation, the so-called rolling motion~\citep{Dupire2012, Lanotte2016}.
Meanwhile, RBCs subjected to high shear rates exhibit the so-called tank-treading motion~\citep{Schmid-Schonbein1969, Fischer1978, Fischer2004}.
The swinging motion was introduced by~\cite{Abkarian2007} as an oscillating orientation of tank-treading motion in the case of relatively low viscosity $\lambda \sim$ 0.5.
Despite these reports,
it is still difficult to control the initial angle of RBCs under flow by experimental techniques.
In the experiment by~\cite{Dupire2012},
they focused more specifically on the cells for which the initial angle of the cell axis of revolution is in the shear plane.
Although previous numerical results showed that a trajectory of RBC orientation angles depend on initial orientations~\citep{Dupire2010},
there is no precise description about bistable flow modes of RBCs under finite $Re$.
The stable configurations of flowing RBCs especially under low $Re$ ($<$ 1) in microchannels smaller than a dozen micrometers have been intensively investigated both experimentally~\citep{Gaehtgen1980, Guckenberger2018, Skalak1969, Tomaiuolo2009} and in numerical simulations~\citep{Freund2011, Freund2014, Guckenberger2018, Takeishi2019Micromach, Takeishi2021}.
For instance,
using a two-dimensional (2D) droplet model,
the numerical studies in~\cite{Kaoui2009} clearly showed that the shape transition in an unbounded Poiseuille flow occurred when a dimensionless vesicle deflation number,
representing shape stability, fell below a certain value.
The effect of degree of confinement $d/W$ (= 0.1-- 0.8, $d$ is the vesicle diameter, $W$ is the channel width) on the vesicle mode was investigated in~\cite{Kaoui2011}.
The three-dimensional (3D) capsule behaviours following neo-Hookean constitutive law in small channels were numerically investigated by~\cite{Hu2012}.
The study revealed that the spherical capsule is more deformed in a circular than in a square cross-section channel even under the same size ratio and flow rate~\citep{Hu2012}.
Those numerical analyses were performed in the Stokes regime.
Although these attempts have revealed velocity-dependent transitions in RBC shapes,
it is still uncertain whether stable RBC shapes in such small microchannels can be reproduced even in larger microchannels.
Furthermore,
much is still unknown in particular about the relationship between the stable flow mode of RBCs,
equilibrium radial position,
and energy expenditure associated with membrane deformations.
To clarify these issues,
we performed numerical simulations of the lateral movement of RBCs under a Newtonian fluid in a circular channel with diameter of $D$ = 50 $\mu$m.
Simulations are performed under a wide range of $Re$ (0.2 $\leq Re \leq$ 30) and $Ca$ (0.05 $\leq Ca \leq$ 1.2), 
as well as with various initial orientation angles $\Psi_0$ and radial positions $r_0/R$.

Our numerical results demonstrate that instead of the parachute and slipper shapes observed in a microchannel with $D$ = 15 $\mu$m~\citep{Takeishi2021},
RBCs have bistable flow modes,
specifically the so-called rolling and tumbling motion,
that depend on the initial orientation angles $\Psi_0$ (figure~\ref{fig:map_ori}).
Most RBCs exhibit a rolling motion (figures~\ref{fig:map_ori}$c$ and \ref{fig:map_ori}$d$),
which is consistent with the results of a previous numerical study of rigid oblate ellipsoidal particles~\citep{Huang2017}.
Especially for low $Ca$ conditions,
rolling RBCs notably appear than tumbling RBCs (figures~\ref{fig:map_ori}$c$ and \ref{fig:map_ori}$d$).
The results are also consistent with previous experimental observations using a circular microcapillary with 50-$\mu$m-diameter~\citep{Lanotte2016},
where the tumbling-to-rolling transition is observed at low shear rates ($\dot{\gamma} <$ 40 s$^{-1}$, corresponding to $Ca$ = 0.05 at $Re$ = 0.2 if the surface shear elastic modulus is considered as $G_s$ = 4 $\mu$N/m).
Our numerical results further show that higher $Re$ conditions impede tumbling motion~(figures~\ref{fig:map_ori}$c$ and \ref{fig:map_ori}$d$).
These modes are associated with different equilibrium radial positions,
where tumbling motions result in more off-centered positions than rolling ones (figure~\ref{fig:re10ca1_2}$c$).
In particular, the equilibrium radial positions of tumbling RBCs can be estimated by spherical capsules (figure~\ref{fig:re10ca1_2_sphere}$b$).
Such equilibrium radial positions are basically independent of initial radial positions $r_0/R$,
except for the case with $r_0/R$ = 0,
where an RBC starts exactly on the channel center ($r_0/R$ = 0) and remains almost on the almost channel axis, $O(\langle r \rangle /R)$ = 10$^{-3}$,
without exhibiting the aforementioned characteristic modes (figure~\ref{fig:re10ca1_2}$c$).
The equilibrium radial coordinates become greater both with decreases in $Ca$ (i.e., stiffer cases) (figure~\ref{fig:effect_ca}$a$) and with increases in $Re$ (figure~\ref{fig:effect_re}$a$).
Stiffer RBCs tend to quickly reach their equilibrium radial positions (figure~\ref{fig:timehist_re10_ca}$b$),
which is qualitatively consistent with a previous numerical study of a spherical hyperelastic particle~\citep{Alghalibi2019}.
On the other hand,
softer RBCs tend to exhibit axial migration,
i.e., inertial migration is impeded by finite deformability (figure~\ref{fig:effect_ca}$a$);
this is qualitatively consistent with previous numerical analyses of spherical capsules in a rectangular channel under Newtonian fluid~\citep{Schaaf2017} and also in polymeric fluid modeled using an Oldroyd-B constitutive equation~\citep{Raffiee2017}.
By simulating multi-spherical-capsule interactions at finite channel $Re$ (= 3--417) in a planar Poiseuille flow considering the SK law,
\cite{Kruger2014} also concluded that the Segr{\'e}-Silberberg effect~\citep{Segre1962} is suppressed upon an increase of the particle deformability.
Considering the effects of initial orientations (figures~\ref{fig:map_ori} and \ref{fig:timehist_re10_ca}$b$) and initial radial positions (figure~\ref{fig:re10ca1_2}$c$) on equilibrium radial positions,
the initial shear stress acting on the cell membrane induces cell deformation and alteration of orientation angles (figures~\ref{fig:map_ori}$a$ and \ref{fig:map_ori}$b$).
The change in cell orientation is primarily affected by the initial orientation angle (figures~\ref{fig:re10ca1_2}$c$),
and the cell subsequently achieves a stable tumbling or rolling motion during radial migration (figure~\ref{fig:timehist_re10_ca}$b$),
with migration toward the equilibrium radial position occurring much later (figure~\ref{fig:timehist_re10_ca}$b$).
Comparison with experimental and numerical results of stable RBC flow mode is our future study.
At the moment, inertial focusing of RBCs in square capillary tube with width of 50 $\mu$m has been well investigated experimentally~\citep{Tanaka2022}.
In the future study, based on this technique, we will confirm aforementioned stable flow mode and mode-depending equilibrium radial position.

To clarify whether the equilibrium radial position minimises the energy expenditure associated with membrane deformations $\langle \delta W_\mathrm{mem}^\ast \rangle$,
the powers are calculated in equation~\eqref{ediss_mem}.
Overall,
off-centered RBCs demonstrate a large velocity gradient (or shear stress),
resulting in large energy dissipation as shown in figures~\ref{fig:effect_ca}($d$), \ref{fig:effect_r0}($d$), and \ref{fig:effect_re}($d$).
The order of magnitude of powers in axially migrated RBCs is relatively small, $O(\langle \delta W_\mathrm{mem} \rangle) \sim$ 10$^{-4}$,
while the powers of off-centered RBCs increase by two orders of magnitudes, $O(\langle \delta W_\mathrm{mem} \rangle) \sim$ 10$^{-2}$ (figures~\ref{fig:effect_r0}$d$, \ref{fig:effect_lam}$c$ and \ref{fig:effect_lam}$d$).
Although the knowledge that large off-centered deformable particles associated with large energy expenditure can be derived by rigid spherical particles,
this tendency is counter to that obtained in a small microchannel with $D$ = 15 $\mu$m and almost no inertia ($Re$ = 0.2)~\citep{Takeishi2021},
where powers $\langle \delta W_\mathrm{mem}^\ast \rangle$ instead decreases as the off-center radial position increases.
The tendency obtained with $\lambda$ = 5 remains the same even at low $\lambda$ (= 1) (figures~\ref{fig:effect_lam}$c$ and \ref{fig:effect_lam}$d$).
The results suggest that aforementioned bistable flow modes in large microchannel ($D$ = 50 $\mu$m) and their equilibrium radial positions cannot be simply determined by the energy expenditure $\langle \delta W_\mathrm{mem}^\ast \rangle$.
The results further show that low $\lambda$ (= 1) conditions impede inertial migration not only for biconcave capsules (non-spherical capsules) but also for spherical capsules (figure~\ref{fig:re10ca1_2_sphere}$b$).
Despite these insights,
we are unsure what factors cause RBCs to adopt a stable shape.
Considering the results observed with powers $\langle \delta W_{\mathrm{mem}}^\ast \rangle$ shown in figures~\ref{fig:effect_lam}($c$) and \ref{fig:effect_lam}($d$),
the stable orientations and equilibrium radial positions of RBCs cannot be explained by the minimum energy dissipation.
The powers $\langle \delta W_{\mathrm{mem}}^\ast \rangle$,
however,
rely on stable flow modes,
equilibrium radial position of RBC centroids,
and viscosity ratios $\lambda$.

Considering that the equilibrium radial positions of rolling RBCs subject to high $Ca$ (= 1.2) increase with $Re \geq$ 15 (figure~\ref{fig:effect_re}a),
a certain level of inertial $Re$ is required in inertial migration of RBCs.
Since the major diameter of rolling RBCs increases by almost 80\% under 15 $\leq Re \leq$ 30 (figure~\ref{fig:effect_re}$b$),
the cell state after inertial migration should be taken into consideration for applications such as cell-sorting techniques.
If the orientation angle of individual RBCs flowing in channels can be manipulated so that the majority of RBCs assume a stable tumbling state,
e.g., by mean of an optical cell rotator~\citep{Kreysing2014},
these tumbling RBCs will accumulate near the wall even under relatively low $Re$ with smaller deformations (figures~\ref{fig:effect_r0}$a$ and \ref{fig:effect_r0}$b$).
Given that the transition can be controlled by adjusting the cell orientation as well as background flow strength,
the results obtained here can be utilised for label-free cell alignment/sorting/separation techniques to precisely diagnose patients with hematologic disorders,
or for the analysis of anticancer drug efficacy in cancer patients.
Our numerical results form a fundamental basis for further studies on cellular flow mechanics.

\section{Conclusion}
We numerically investigate the lateral movement of RBCs with a major diameter of 8 $\mu$m under a Newtonian fluid in a circular channel with 50-$\mu$m diameter.
Simulations are performed for a wide range of $Re$ and $Ca$, 
as well as various initial orientation angles and radial positions.
The RBCs are modelled as a biconcave capsule,
whose membrane follows the SK law.
The problem is solved by the LBM for the inner and outer fluids,
and the FEM is used to follow the deformation of the RBC membrane.
The numerical results show that RBCs have bistable flow modes,
the so-called rolling motion and tumbling motion,
which depend on the initial cell orientations and which are established soon after flow onset.
The vast majority of RBCs exhibit the rolling motion.
Furthermore, higher $Re$ conditions impede tumbling motion.
These modes are associated with different equilibrium radial positions,
with tumbling RBCs flowing much further away from the channel axis than rolling ones.
RBCs subject to high $Ca$ (i.e., large deformability) tend to exhibit axial migration even for finite $Re$,
but inertial migrations are enhanced over a certain value of $Re$.
The inertial migration of RBCs involves the alternation of orientation angles,
which are primarily affected by the initial orientation angles.
The RBCs then adopt the aforementioned bistable modes during the migration,
followed by further migration to the equilibrium radial position at much later time periods.
The stable orientations and equilibrium radial positions of RBC centroids do not minimises the energy expenditure associated with membrane deformations.
The energy expenditure, however, rely on stable flow modes, the equilibrium radial positions of RBCs, and viscosity ratios.

\section*{Acknowledgements}
This research was supported by JSPS KAKENHI Grant Numbers JP20H04504 and JP20H02072,
and by the Keihanshin Consortium for Fostering the Next Generation of Global Leaders in Research (K-CONNEX),
established by the Human Resource Development Program for Science and Technology.
N.T. is grateful for the financial support of UCL-Osaka Partner Funds.

\section*{Conflicts of Interest}
The authors report no conflict of interest.

\section*{Supplementary movie}
Supplementary movies are available at \textit{https://doi.org/xxx.yyy.zzz}.

\appendix

\section{Numerical setup}\label{appA3}
\begin{figure}
  \centering
  \includegraphics[height=5.5cm]{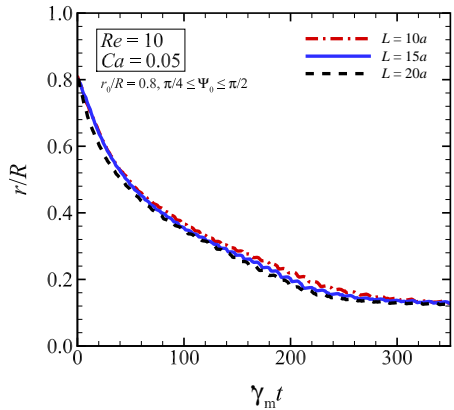}
  \caption{
	  Time history of the radial position $r/R$ of an RBC centroid with $D$ = 50 $\mu$m for different channel lengths $L$ (= 10$a_0$, 15$a_0$, and 20$a_0$).
	  The results are obtained with $Re$ = 10,
	  $Ca$ = 0.05,
	  $\upi/4 \leq \Psi_0 \leq \upi/2$,
	  and $r_0/R$ = 0.8.
  }
  \label{fig:effect_length}
\end{figure}

We have tested the channel length $L$,
and investigated its effect on the radial positions of RBC centroids.
Although previous numerical studies of deformable particles have set a variety of computational lengths depending on $Re$, e.g.~\citep{Alghalibi2019, Kilimnik2011, Raffiee2017, Schaaf2017},
we tested the trajectory of the radial positions of RBC centroids for different channel lengths $L$ (= 10$a_0$, 15$a_0$, and 20$a_0$).
The results of the time history of the radial position of RBC centroid $r$ are compared between these different channel lengths in figure~\ref{fig:effect_length},
where the centroid position $r$ is normalised by the channel radius $R$.
The results are consistent among all cases,
and hence the results presented in this study are all obtained with the channel length of $L = 20a$.

\begin{figure}
  \centering
  \includegraphics[height=5.5cm]{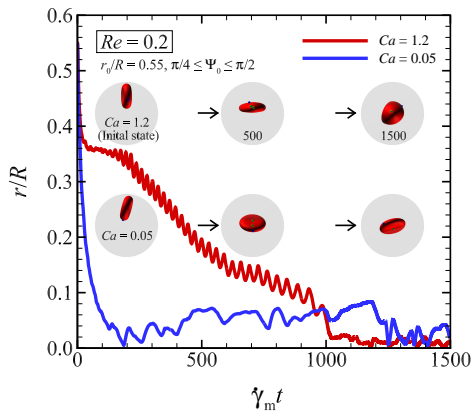}
  \caption{
	  Time history of the radial position $r/R$ of an RBC centroid with $D$ = 20 $\mu$m for different $Ca$ (= 0.05 and 1.2).
	  The results are obtained with $Re$ = 0.2,
  	  $\upi/4 \leq \Psi_0 \leq \upi/2$,
	  and $r_0/R$ = 0.55.
	  The insets represent snapshots of cross-sectional area of RBCs subject to each $Ca$ at specific times:
	  $\dot{\gamma}_\mathrm{m}t$ = 0 (initial state), 500, and 1500, respectively. 
  }
  \label{fig:d20re02}
\end{figure}

We have also confirmed the axial migration for a relatively small channel with $D$ = 20 $\mu$m (i.e., $d/D$ = 0.2),
where the channel length remains the same as $L$ = 20$a_0$. 
Figure~\ref{fig:d20re02} shows the time history of $r/R$ for different $Ca$ (= 0.05 and 1.2).
Although a highly deformable RBC (i.e., large $Ca$ = 1.2) takes much longer to migrate toward the channel center than a stiffer RBC (i.e., small $Ca$ = 0.05),
both RBCs finally complete the axial migration.
This result also indicates that the axial migration of RBCs will occur even in high $Ca$ (= 1.2) at least for $d/D \leq$ 0.2,
instead of off-centered slipper shapes of RBCs at $d/D \approx$ 0.53 ($d$ = 8 $\mu$m and $D$ = 15 $\mu$m)~\citep{Takeishi2021}.

\section{Membrane mechanics}\label{appA2}
Since the RBC membrane is very thin relative to its major diameter,
we can consider the deformation of its median surface in the absence of bending resistance.
Furthermore, the stress can be integrated across the thickness and be replaced by tensions, i.e., forces per unit length.
Consider a material point on the surface of a two-dimensional membrane.
Let $\v{x}$ be the position of the material point in a deformed state.
In fixed Cartesian coordinates, it is defined as
\begin{equation}
	\v{x} = x^i \v{e}_i 
  \label{x}
\end{equation}
where $\v{e}_i (i \in [1,3])$ is the Cartesian basis. 
We also introduce local curvilinear coordinates on the membrane $(\xi^1, \xi^2, \xi^3)$,
and the local covariant bases are defined by
\begin{align}
	&
	\v{G}_1 = \frac{\partial \v{X}}{\partial \xi^1}, \
	\v{G}_2 = \frac{\partial \v{X}}{\partial \xi^1}, \
	\v{G}_3 = \v{n}, \\ 
	&
	\v{g}_1 = \frac{\partial \v{x}}{\partial \xi^1}, \
	\v{g}_2 = \frac{\partial \v{x}}{\partial \xi^1}, \
	\v{g}_3 = \v{n}, 
  \label{G_g}
\end{align}
where $\v{X}$ and $\v{x}$ are material points on the membrane in the reference and deformed state, respectively.
$\v{n}$ is the unit normal outward vector,
which is calculated as
\begin{align}
	\v{g}_3
	= \frac{\v{g}_1 \times \v{g}_2}{|\v{g}_1 \times \v{g}_2|}
	= \v{g}^3
	= \frac{\v{g}^1 \times \v{g}^2}{|\v{g}^1 \times \v{g}^2|}
	= \v{n}.
\end{align}
The associated contravariant bases are defined as $\v{g}^i \cdot \v{g}_j = \delta_j^i$, 
where $\delta$ is the Kronecker's delta.
The covariant and contravariant metric tensors can be written as
\begin{equation}
	g_{ij} = g_{ji} = \v{g}_i \cdot \v{g}_j, \quad
	g^{ij} = g^{ji} = \v{g}^i \cdot \v{g}^j,
\end{equation}
where $g_{i3} = \delta_{i3}$ and $g^{i3} = \delta^{i3}$.
The local, in-plane deformation of the membrane can be measured by the Green-Lagrange strain tensor $\v{E}$
\begin{equation}
	\v{E} = \frac{1}{2} \left( \v{g}_{\alpha\beta} - \v{G}_{\alpha\beta} \right),
	\quad (\alpha, \beta \in [1, 2]).
  \label{E}
\end{equation}
The two invariants of the strain tensor are given by 
\begin{equation}
	I_1 = g_{\alpha\beta}G^{\alpha\beta} - 2, \quad
	I_2 = |g_{\alpha\beta}||G^{\alpha\beta}| - 1,
  \label{I1_I2}
\end{equation}
where $|g_{\alpha\beta}| (= g_{11}g_{22} - g_{12}g_{21})$ is the determinant of the metric tensor (similarly for the reference state, $|G_{\alpha\beta}|$).
The contravariant expression of the Cauchy tensor $\v{T}$ is then given by
\begin{equation}
	\v{T} = \frac{2}{J_s} \frac{\partial w}{\partial I_1} G^{\alpha\beta} + 2 J_s \frac{\partial w}{\partial I_2} g^{\alpha\beta},
  \label{T}
\end{equation}
where $w$ is the surface strain energy function,
and $J_s$ is the Jacobian,
which expresses the area dilation ratio.
In this study,
the SK law (equation~\ref{SK}) is considered for $w$.

\bibliographystyle{jfm}
\bibliography{jfm-instructions}

\end{document}